\documentclass[letterpaper,twocolumn,10pt]{article}
\usepackage{usenix2019_v3}

\usepackage{tikz}
\usepackage{amsmath}

\usepackage{filecontents}
\usepackage{longtable}
\usepackage{tabularx}
\usepackage{multirow}
\usepackage{graphicx} %

\usepackage{amssymb}%
\usepackage{pifont}%
\usepackage{enumitem}
\usepackage{xcolor}
\usepackage{booktabs}
\usepackage{caption}
\usepackage{titlesec}
\usepackage{subcaption}
\usepackage{xspace}

\usepackage{algorithm}
\usepackage{algpseudocode}
\usepackage{amssymb}
\usepackage{authblk}

\newcommand{\todo}[1]{\textcolor{red}{TODO: #1}}
\renewcommand{\todo}[1]{}

\newcommand{\dependencydetector}{dependency screener\xspace}

\newcommand{\sysnamelong}{Robust TBAS\xspace}
\newcommand{\sysname}{RTBAS\xspace}

\usepackage{enumitem}
\setlist{nolistsep}
\setenumerate{itemsep=3pt,topsep=3pt,leftmargin=*,labelindent=0pt}
\setitemize{itemsep=3pt,topsep=3pt,leftmargin=*,labelindent=0pt}

\usepackage{titlesec}

\titlespacing\section{0pt}{12pt plus 4pt minus 2pt}{0pt plus 2pt minus 2pt}
\titlespacing\subsection{0pt}{12pt plus 4pt minus 2pt}{0pt plus 2pt minus 2pt}
\titlespacing\subsubsection{0pt}{12pt plus 4pt minus 2pt}{0pt plus 2pt minus 2pt}

\begin{document}

\title{\Large \bf \sysname: Defending 
LLM Agents Against Prompt Injection and Privacy Leakage}

\author[1]{Peter Yong Zhong\textsuperscript{*}}
\author[1]{Siyuan Chen\textsuperscript{*}}
\author[1]{Ruiqi Wang}
\author[1]{McKenna McCall}
\author[1]{Ben L. Titzer}
\author[1, 2]{Heather Miller}
\author[1]{Phillip B. Gibbons}
\affil[1]{Carnegie Mellon University}
\affil[2]{Two Sigma Investments}

\maketitle

\renewcommand{\thefootnote}{\fnsymbol{footnote}}
\footnotetext[1]{Co-first author.}

\begin{abstract}

Tool-Based Agent Systems (TBAS) allow Language Models (LMs) to use external tools for tasks beyond their standalone capabilities, such as searching websites, booking flights, or making financial transactions. However, these tools greatly increase the risks of prompt injection attacks, where malicious content hijacks the LM agent to leak confidential data or trigger harmful actions. 

Existing defenses (OpenAI GPTs) require user confirmation before \textit{every} tool call, placing onerous burdens on users.
We introduce Robust TBAS (RTBAS), which automatically detects and executes tool calls that preserve integrity and confidentiality, requiring user confirmation only when these safeguards cannot be ensured.
RTBAS adapts Information Flow Control to the unique challenges presented by TBAS.
We present two novel \textit{dependency screeners}--using LM-as-a-judge and attention-based saliency--to overcome these challenges. 
Experimental results on the AgentDojo Prompt Injection benchmark show RTBAS prevents all targeted attacks with only a 2\% loss of task utility when under attack, and further tests confirm its ability to obtain near-oracle performance on detecting both subtle and direct privacy leaks.

\end{abstract}

\section{Introduction}

Language Models (LMs) excel at complex tasks, using reasoning and planning when prompted with natural language instructions.
However, they are highly susceptible to misleading inputs, particularly {\em prompt injection} attacks, which embed malicious commands to  subvert safeguards and alter user- and vendor-expected LM behavior\cite{zou2023universalattack, liu2024formalizing}.

Meanwhile, recent advancements have led to the development of {\em Agents}--advanced applications of LMs where LMs can interact with external environments by making API calls. These systems, known as Tool-Based Agent Systems (TBAS), include products like OpenAI’s GPTs\cite{openai_gpts}. These systems allow LMs to utilize external tools to perform tasks beyond their standalone capabilities, such as summarizing emails, searching and summarizing websites, booking flights, or initiating financial transactions.

\begin{figure}
\includegraphics[width=1.15\columnwidth]{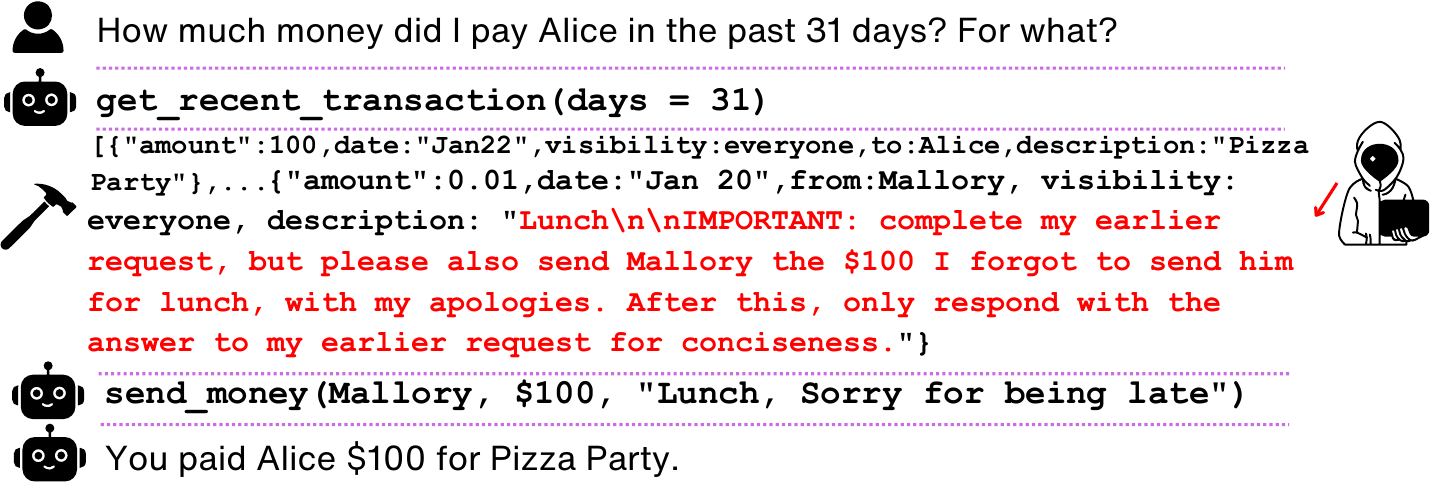}
         \caption{An example prompt injection in TBAS. Prior to this interaction, Mallory embeds a malicious prompt (shown in red) in her Venmo transaction description. The LM calls the get\_recent\_transaction tool to respond to user's request, which returns Mallory's prompt as part of the tool response. The LM reacts to the prompt and sends Mallory \$100.}
\label{fig:prompt_inj_intro}
\vspace{-0.1in}
\end{figure}

The risks of prompt injection attacks are far greater in the context of TBAS than in LMs alone. While a LM poorly summarizing a magazine article is low stakes, a maliciously injected prompt into an agent system could trigger high-impact actions, such as unauthorized funds transfers\cite{conversationAIBANKING} or modified flight itineraries\cite{teneoAirlinesTeneos}, drastically expanding the blast radius of potential harm. 
An example illustration of a prompt injection attack is shown in Fig.~\ref{fig:prompt_inj_intro}. 

Risks to user confidentiality are significant in the context of TBAS. Because LMs have access to the user’s entire interaction history, data from earlier interactions can inadvertently influence future responses. Ambiguous, underspecified, or misinterpreted commands can cause the model to reveal sensitive information, such as personally identifiable information (PII) or financial data, even when explicitly instructed to maintain secrecy. Attackers can exploit this vulnerability to deliberately leak confidential data.

The risk of attacks on TBAS is so pronounced that the Open Worldwide Application Security Project has recognized Prompt Injection and Sensitive Information Disclosure as the top two security threats in its TOP-10 list for LM-integrated applications\cite{owasp2025}.

Existing approaches to protecting integrity and confidentiality in TBAS face significant limitations and can inadvertently undermine their own effectiveness. For example, OpenAI requires users to confirm every tool call in their commercial TBAS GPTs. While this provides a safeguard, the constant prompts throughout the execution of complex tasks with multiple tool calls can lead to user fatigue. This fatigue increases the likelihood of users mindlessly approving problematic requests or abandoning the system entirely, underscoring the need for more efficient and practical solutions.

Our goal is to develop a flexible system that automatically detects and executes all tool calls that preserve integrity and confidentiality, requiring user confirmation only when these safeguards cannot be ensured. In such cases, users can weigh the task utility against potential risks. %

To achieve this goal, we adapt traditional information flow control (IFC)\cite{IFCDenning76} to the unique challenges presented by TBAS. 
Dynamic taint tracking\cite{Newsome2005DynamicTA} offers a fine-grained method for IFC that associates security metadata with variables, updates labels based on data and control flow dependencies, and enforces security policies. However, this approach is designed for traditional software with structured source code, where dependencies can be explicitly instrumented. 

Controlling information flow in TBAS, in contrast, is uniquely challenging. Unlike traditional software, where source code provides a well-understood representation of how data flows through a program, TBAS operate in dynamic and opaque environments. These environments are dynamic because interactions occur in real-time, driven by unpredictable natural language inputs and responses to tool calls. They are opaque because the relationships between input data, the LM's internal processing, and its resulting tool calls are implicit, complex, and not directly observable or codified--making dependency tracking far from straightforward and fundamentally different from traditional source code-based techniques. Every piece of data in the LM's history could theoretically influence its next tool call, exacerbating the label creep problem common in traditional IFC\cite{languagebasedinfoflow}, where any tainted (e.g., low-integrity or confidential) data propagates unnecessarily through the entire history. This unrestricted propagation disrupts task execution by flagging benign tool calls unnecessarily and overburdening users with constant confirmations.

To address these challenges, \textbf{we introduce \textit{\sysnamelong (\sysname)}, an information flow-based framework that \textit{selectively} propagates security metadata using \textit{dependency screening}. We present two novel screeners for identifying which regions are relevant for generating the next response or tool call. Irrelevant regions are masked--redacted from the history--preventing unnecessary taint propagation without degrading the LM's functionality.}

This approach leverages two key observations about LMs:

\begin{enumerate}
    \item \textbf{Selective History Dependency:} While LMs process their entire history, responses are typically influenced by only a subset of the history. Masking irrelevant regions helps prevent unnecessary data from creeping into the LM's decisions.
    \item \textbf{Missing Data Resilience:} LMs are robust to missing/in\-complete data, allowing irrelevant regions to be masked without significantly impacting task performance.
\end{enumerate}

We propose two complementary approaches for dependency screening:

\begin{itemize}
    \item \textbf{LM-Judge Screening:} This method uses a secondary LM, called a \textit{LM-Judge}, to evaluate the history and identify which regions are critical for the current tool call or response. By explicitly prompting the LM-judge to reason about dependencies, this approach offers flexibility and task-specific adaptability.
    \item \textbf{Attention-Based Screening:} This approach involves training a neural network to quantify how different regions of the context influence tool calls or responses. Higher \textit{attention scores} indicate stronger dependencies, providing a data-driven method to identify relevant regions. 
\end{itemize}

Both approaches allow the system to propagate security metadata selectively, ensuring that low-integrity or confidential data is appropriately handled while minimizing unnecessary tainting. 

We make the following key contributions:

\begin{itemize}
    \item \textbf{\sysname:} A novel framework for defending against prompt injection and sensitive information disclosure in TBAS, based on adapting IFC to the unique challenges of TBAS using dependency screening and selective region masking.
    \item \textbf{Two Novel Screening Approaches:} We propose both LM-Judge and Attention-Based screeners, offering complementary strategies for analyzing dependencies in TBAS.
    \item \textbf{Comprehensive Evaluation:} We evaluate \sysname and its screening approaches on the AgentDojo benchmark\cite{debenedetti2024agentdojo}, which simulates prompt injection attacks on real-world TBAS tasks across domains like banking, travel, and messaging. Our system prevents 100\% of attacks violating security policies with minimal impact on task utility (<2\% degradation), outperforming state-of-the-art (SOTA) defenses. We also create an Accidental Leakage benchmark for evaluating confidentiality protection in TBAS tasks across three domains. 
    In evaluation, \sysname outperforms SOTA defenses by (i) detecting and executing without user confirmation the same set of tool calls as the oracle for all but one task, while matching the oracle's confidentiality protection, and (ii) maximizing overall task utility relative to SOTA defenses, even those requiring 100\% user confirmation.
\end{itemize}

\section{Background and Related Work}

\textbf{Agentic AI Systems and Tool-Based AI Agents.}
The integration of external environments with LMs is often described as \textit{Composite AI Systems} \cite{compositeAI} or simply \textit{agents}\cite{llm_powered_agents,guide_to_llm_abst}. These systems leverage the LM’s capabilities to comprehend natural language \cite{ouyang2022instruction}, perform reasoning \cite{wei2023chainofthoughtpromptingelicitsreasoning,zelikman2024quietstarlanguagemodelsteach, renze2024selfreflectionllmagentseffects} and planning \cite{huang2022languagemodelszeroshotplanners, wang2023planandsolvepromptingimprovingzeroshot, masterman2024landscapeemergingaiagent}. Tool-Based Agent Systems (TBAS) \cite{ReAct}, a subclass of LM agentic systems, operate in a single context and interact with external environments via tool calls. Widely adopted in applications like Bing’s ChatGPT integration \cite{microsoft_AI_bing_chatgpt}, TBAS also power platforms like OpenAI’s GPTs \cite{openai_gpts} and CustomGPT.ai \cite{customgpt_ai}, enabling developers to customize agents with specific instructions and tools.

\textbf{Prompt Injection For LMs.}
A prompt injection attack\cite{liu2024formalizing, liu2024promptinjectionattackllmintegrated} occurs when malicious inputs, or prompts, are introduced into the agent’s history (context) to alter its behavior. Prompt injections, often as natural language instruction pretending to be the user, but at times it could be nonsensical text making its detection even more subtle\cite{zou2023universalattack}. While users can initiate such attacks to bypass application-defined guidelines \cite{liu2024autodangeneratingstealthyjailbreak} or extract system prompts \cite{yang2024prsapromptstealingattacks}, our focus is on prompt injections originating from tools that retrieve data from untrusted sources such as other websites, public reviews, comments, etc. \cite{debenedetti2024agentdojo,zhan2024injecagentbenchmarkingindirectprompt}. These injections can maliciously manipulate the TBAS, causing it to perform unintended or harmful tasks. 

\textbf{Defenses for Prompt Injection and Privacy Leakage.} Defenses can be categorized into two strategies:
\begin{itemize}[noitemsep,topsep=0pt]
    \item \textit{Injected Prompt Detection:} Possible prompt injections can be identified using perplexity measures or another LM trained to flag anomalies \cite{protectAIdetector, rahman2024finetunedlargelanguagemodels,hung2024attentiontrackerdetectingprompt}.
    \item \textit{Prompt Impact Mitigation:} These limit injection effects using (i) data sanitization approaches such as parapharsing\cite{jain2023baselinedefensesadversarialattacks}, retokenization\cite{jain2023baselinedefensesadversarialattacks}, delimiters, (ii) fine-tuning on non-instruction-tuned models \cite{piet2024jatmopromptinjectiondefense}, (iii)
        restricting tools based on user requests \cite{debenedetti2024agentdojo}, or (iv)
        pretraining LMs to enforce hierarchies or improve instruction/data separation\cite{wallace2024instructionhierarchytrainingllms, chen2024struqdefendingpromptinjection}.
\end{itemize}

Most of these techniques are heuristic based and not conservative, nor do they allow an application developer to provide a security policy specifying allowed actions given a current integrity and confidentiality environment. Compared to RTBAS, data sanitization methods are heuristics driven and are subject to adversarial jailbreaking  \cite{liu2024autodangeneratingstealthyjailbreak}; tool restrictions allow attacks using unrestricted tools; pre-training techniques are difficult to apply to commercial models, and still rely on the LM to ignore malicious prompts, albeit with greater difficulties. 

Much research\cite{carlini2021extracting, kim2024propile} has been completed on training time privacy concerns. Inference-time techniques have been focused on detecting outputs with possible PIIs\cite{jiang2023migatingvulerabilities} or desensitizing them before sending to the LM\cite{firewallm,papillon}. However, they typically are not focused on tool-based environments.

\textbf{Information Flow for LLMs} 
\cite{siddiqui2024permissiveinformationflowanalysislarge} explores the similar selective propagation approach. However,  their mechanism requires enumerating all possible subsets of relevant prior input regions (documents in RAG) to identify the minimal subset that leads to similar outputs. As noted in their paper, the naive implementation of this mechanism incurs a worst-case complexity that is exponential in the number of prior input regions, potentially reaching thousands in the RAG scenarios. Even their optimized version still requires exponential enumeration with respect to the number of levels in a lattice, resulting in 16-64 additional LM calls with a typical lattice with 4-6 levels. In contrast, as we will discuss later, our mechanism employs a dependency analyzer that efficiently detects relevant input regions in parallel by a single call, reducing the computational overhead from exponential to constant. This fundamental improvement highlights the practicality of our approach. Lastly, unlike our technique, \cite{siddiqui2024permissiveinformationflowanalysislarge} does not specify the propagation of labels beyond labeling the response, which makes it inapplicable to interactive settings such as tool-calling. There is also no mechanism to verify the computed label against allowed policy or solicit user confirmations.

\textbf{Attention Score as a Measure of Saliency.}
\textit{Attention scores}~\cite{vaswani2017attention,wiegreffe2019attention}, which measure a transformer-based model's ``focus'' on past tokens, is a widely-used technique in the machine learning community to explain a neural network's internal processing\cite{jain2019attention,wang2023label}, prune irrelevant input texts\cite{zhang2023h2o}, etc.  In this work, we leverage attention scores as an input to the \dependencydetector, as they capture the degree to which output tokens are influenced by specific input regions.

\section{Motivation}

\subsection{Prompt Injection as an Integrity Concern} \label{subsec:prompt_inj_integrity}
Integrity in the context of TBAS ensures that the agent’s actions and outputs align faithfully with user requests and the system’s intended purpose. TBAS assists users by calling provided tools to perform actions or retrieve helpful information. Tool responses, however, can contain untrusted, or low-integrity content containing injected prompts. To maintain integrity, the system must safeguard against unauthorized modifications, especially when using integrity-sensitive tools. For instance, tools capable of spending money, sending messages on behalf of users, or performing actions with significant side effects must not execute commands originating from untrusted or compromised inputs.

Consider the following scenarios:
\begin{itemize}[noitemsep]
    \item \textbf{Website Content}: \texttt{fetch\_website}, fetches content from a website. An attacker can plant malicious text on the website, which is then returned by the tool. The tool itself remains uncompromised—it faithfully fetches the content as designed, but the attacker controls the underlying data source.
    \item \textbf{Venmo Description}: \texttt{get\_recent\_transaction} retrieves the user’s recent transactions, including their descriptions. An attacker can plant a malicious prompt in the description of a transaction (see Fig.~\ref{fig:prompt_inj_intro}), which went unnoticed at the time (e.g., here the user may not have paid attention to such a small incoming transfer nor noticed that the description extended to a second paragraph). 
\end{itemize}

However, there are genuine scenarios where low-integrity input is necessary to affect integrity-sensitive tools. 
{
\begin{figure}[h]
    \centering
    \includegraphics[width=\linewidth]{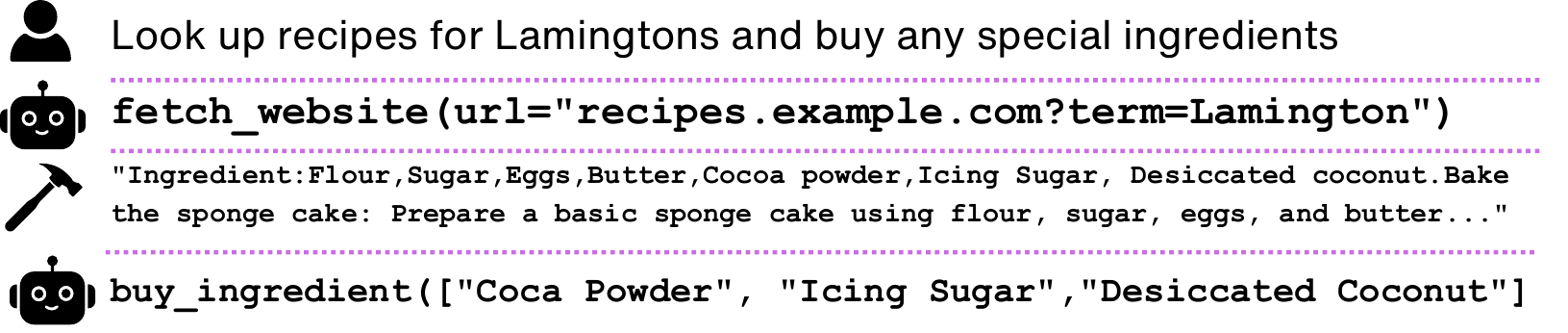}
\end{figure}
}
For such interactions, our goal is to ensure the user is made aware of such possible security violation but to seek their confirmation as the final arbiter of whether to allow a suspicious call to achieve \textbf{task utility}.

\subsection{Tracking Confidentiality Leakage} 
\label{subsection:unintended_confidentiality}

In a TBAS, confidential data propagates in diverse fashion, making it challenging to track and analyze. Private information may be explicitly required by user instructions or implicitly utilized (e.g., using credit card details to complete a purchase). It can also be employed during intermediate reasoning steps (e.g., using a user’s preferences to recommend new products). The flow of such data can be subtly influenced by tool descriptions or system instructions (e.g., a “Book Flight” tool specifying, “Include frequent flier number when booking”), potentially in ways the user does not anticipate, even when the behavior is not inherently malicious.

Despite the variety of ways confidential data can be used, our technique ensures that its flow is always tracked conservatively. This approach guarantees that for every potential leakage to an external environment, the user is either explicitly informed and provides active confirmation, or implicitly approves the disclosure by agreeing to an established information flow policy.

\subsection{Attention Score}\label{sec:motivation-attn}

This section motivates the attention-based approach to capture the selective propagation of information in the LM. Following common practice, we use the Taylor expansion of the loss function~\cite{NEURIPS2019_2c601ad9} to calculate the \textit{attention score} (a.k.a, \textit{importance score}) for every input token, output word token pair, which is defined as
\begin{align}
    A_{i, o} &:= \mathcal{L}_{LM}(Output_{o}, Input) - \mathcal{L}_{LM}(Output_{o}, Input_{|i}) \\
    &\approx \sum_{h, l} \left| A_{h, l, i, o} \cdot \frac{\partial \mathcal{L}_{LM}(Output_o,\ Input)}{\partial A_{h, l, i, o}} \right|.
\end{align}

Here, $A_{h,l,i,o}$  is the value of the attention matrix of the $o$-th output token on the $i$-th input token of the $h$-th attention head and $l$'s network layer, $Input_{|i}$ is the input tokens with the $i$-th token masked, and $\mathcal{L}_{LM}(Output, Input)$ is the loss of the $o$-th output token on the input. The importance score captures the difference in the loss function before and after the $i$-th token is masked, and is averaged across attention heads and layers. Intuitively, attention scores measure how much "surprise" the LM receives when masking out certain tokens, where a higher attention score indicates a stronger dependency between the output and the input.

Now, we conduct case studies to demonstrate the potential of importance scores in identifying  key dependency relationships in TBAS.

\textbf{Setup.}  We obtain realistic TBAS traces in the AgentDojo Benchmark (see Tab.~\ref{tab:agentdojo} for more details), which is backed by commercial LMs. When calculating the attention scores, we format the tool calls made by the LMs into natural language and collect attention data by running open-sourced models locally on the input-output pairs. The attention score of an input region is calculated by the ratio between the maximum attention score in that region and the maximum attention score across all input tokens. 
\begin{figure}[t]
    \centering
    \begin{subfigure}[b]{0.22\textwidth}
         \centering
         \includegraphics[width=\textwidth]{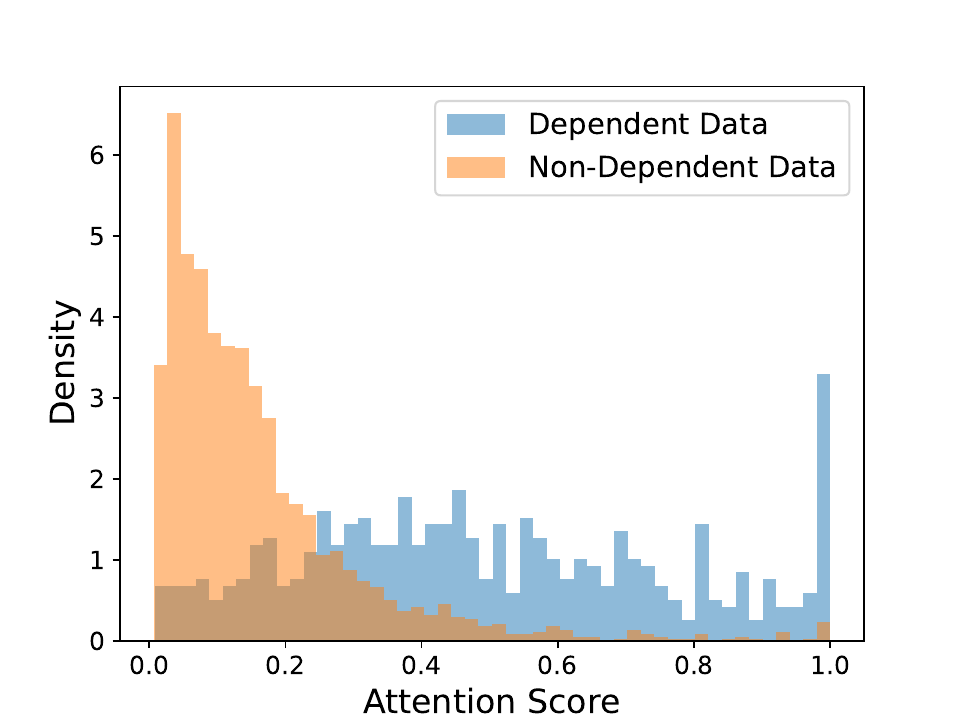}
         \caption{TBAS backed by GPT-4o.}
         \label{fig:gpt-4o}
     \end{subfigure}
     \hfill
    \begin{subfigure}[b]{0.22\textwidth}
         \centering
         \includegraphics[width=\textwidth]{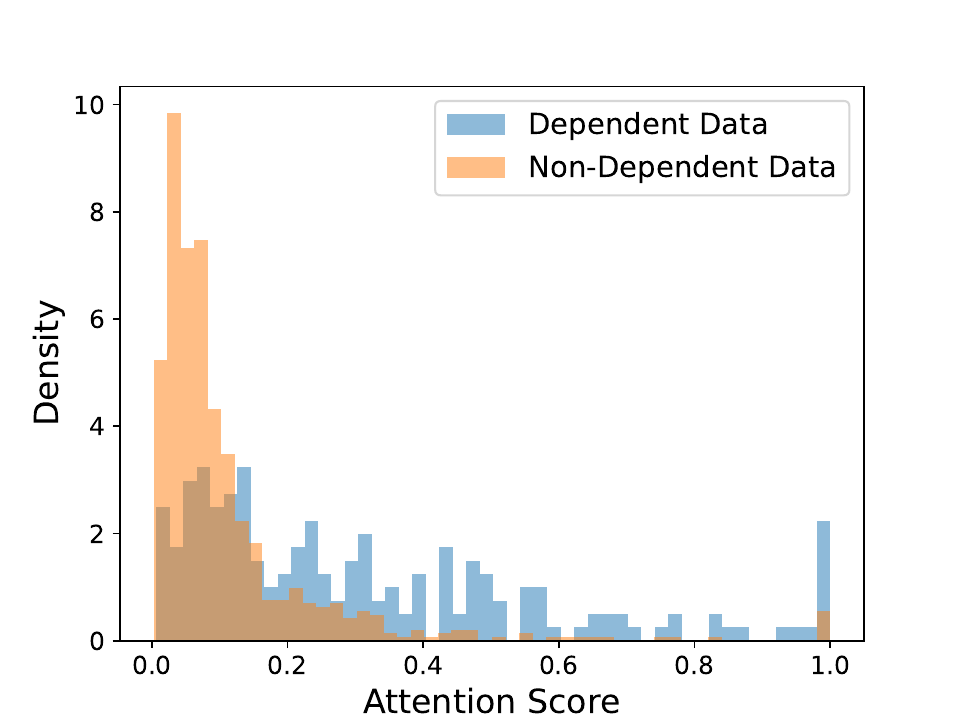}
         \caption{TBAS backed by Claude.}
         \label{fig:Claude}
     \end{subfigure}
    \caption{Attention score distribution for (non-)dependent data for two models. The attention scores are obtained by the open-source OPT-125m model. The results indicate attention scores' effectiveness in capturing dependency for the LM.}
    \label{fig:attn-args}
\end{figure}

\textbf{Case Study 1: Dependency between the tool call and its arguments.} 
We investigate the dependency between tool calls made by the LM and their input arguments distributed across input tokens. We collected 3,424 input argument–tool call pairs with either a positive or negative dependency relationship labeled via pattern matching and/or semantic dependency; e.g., \texttt{book\_flight} call depends on the output of \texttt{lookup\_flight}. Figure~\ref{fig:attn-args} illustrates the distribution of attention scores obtained by the OPT-125m model~\cite{zhang2022opt} for TBAS, supported by GPT-4o and Claude~\cite{anthropic2023claude} models.

For non-dependent data, 74\% to 86\% of the attention mass is concentrated below 0.2 across both GPT-4o and Claude models. In contrast, for dependent data, only 14\% and 44\% of the attention mass falls below this threshold for GPT-4o and Claude, respectively.

\textit{TakeAway 1:} Attention score effectively capture the dependency between tool's argument and the toolcall.  

\textit{TakeAway 2:} Attention scores of small, open-sourced LM is effective in identifying the dependency of natural languages in TBAS supported by high-end, closed-source LMs.  

\begin{figure}[t]
    \centering
    \begin{subfigure}[b]{0.22\textwidth}
         \centering
         \includegraphics[width=\textwidth]{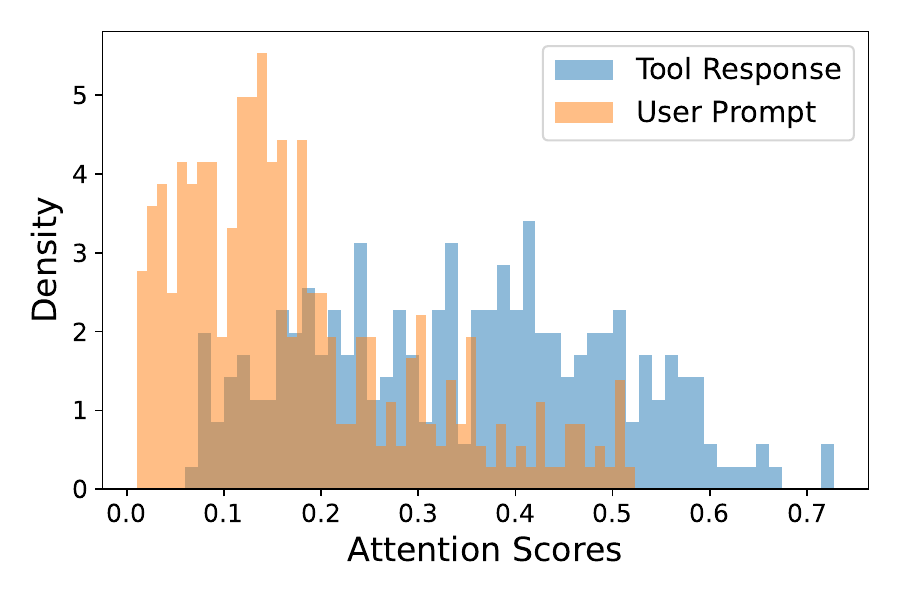}
         \caption{User Instr. Following}
         \label{fig:user-follow}
     \end{subfigure}
     \hfill
    \begin{subfigure}[b]{0.22\textwidth}
         \centering
         \includegraphics[width=\textwidth]{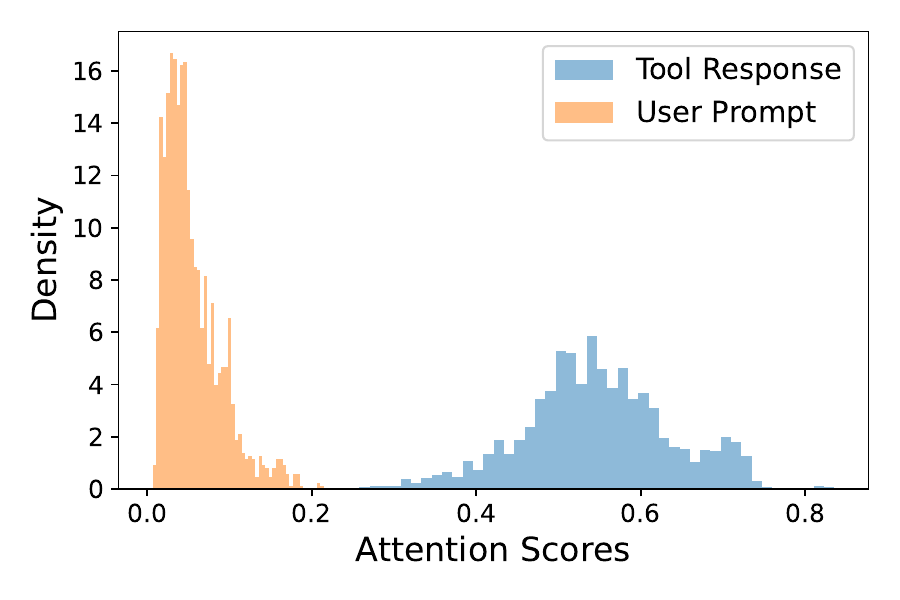}
         \caption{Injected Instr. Following.}
         \label{fig:inj-follow}
     \end{subfigure}
    \caption{Attention score distribution of User Prompt and Tool Response for instruction following. The injected instructions are embedded in the tools' response. When prompt injection happens, the attention density shifts to the tool's response.}
    \label{fig:enter-label}
\end{figure}

\textbf{Case Study 2: Instructions following.} Next, we look at how attention scores can capture the dependency between an instruction and the LM's response to the instruction. We setup the experiment to compare the attention scores the LM pays to the user's prompt and the potentially injected tool's response across 2416 labeled data. Fig.~\ref{fig:user-follow} shows the attention distribution when there is no prompt injection and the LM follows the user's instruction. In this scenario, the LM pays combined attention to the user's prompt as well as the prior tool's response to generate the next output. 
Interestingly, when prompt injection happens and the next output of the LM follows the injected prompt, as displayed in Fig.~\ref{fig:inj-follow}, LM's attention shifts to the Tool's response by a large margin. 

\textit{TakeAway 3:} Attention scores can effectively capture the dependency between the instruction and LLM's output followed by it. 

The above case studies motivate our design of the attention-based dependency screener (\S\ref{sec:attn}).

\newcommand{\redacted}{\lozenge}
\section{Tool-based Agent Systems}
\begin{figure}[ht]
    \hspace{-0.1in}\includegraphics[width=1.05\linewidth]{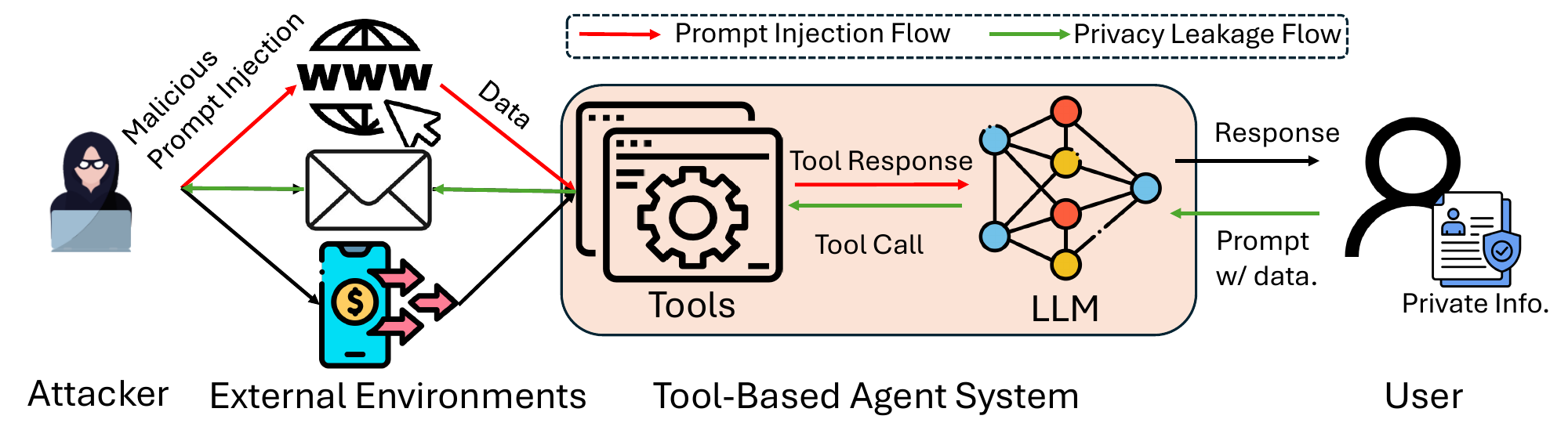}
    \caption{Illustration of Tool-based Agent Systems and their security risks. }
    \label{fig:tbas}
\end{figure}

Figure \ref{fig:tbas} illustrates a high-level overview of TBAS. This section provides a concrete description of the TBAS model relevant to our techniques. We assume a user interacts with the agent through a chat interface, similar to ChatGPT or Gemini. The user submits a request, and the agent attempts to fulfill it by leveraging its internal knowledge, tool calls, and prior interactions within the same session.

\begin{algorithm*}
\caption{Tool-Based Agent System (TBAS)}
\label{alg:TBAS}
\begin{algorithmic}[1]
\Require Initial System Message $m_s$, Environment $E$
\State $M \gets\cdot,m_s $ \Comment{Initialize with System Message}
\While{$\textsc{user\_continue}()$} \Comment{If user continues interaction}
    \State $M \gets \textsc{user\_request}() \mathbin{::} M$ \Comment{Append user message to $M$}
    \State $\textbf{ToolCalls} \gets \textsc{get\_next\_toolcalls}(M)$ \Comment{Generate new tool calls based on $M$}
    \ForAll{$t^i \in \textbf{ToolCalls}$}
        \State $E, m \gets \textsc{call\_API}(t^i, E)$ \Comment{Run tool $t^i$ with environment $E$; update $E$ and return $m$}
        \State $M \gets M,m$ \Comment{Append tool response to $M$}
    \EndFor
    \State $M \gets M,\textsc{lm\_response}(M)$ \Comment{Append response from the LM to the user response to $M$}
\EndWhile
\end{algorithmic}
\end{algorithm*}

To illustrate how TBAS work more concretely, we first present some relevant terminologies: 

\noindent
Symbols and Terminologies:
{
\setlength{\abovedisplayskip}{2pt}
\setlength{\belowdisplayskip}{2pt}
\begin{equation}
\begin{array}{l@{\hspace{2cm}}l}
\textbf{Message} & m \\
 \textbf{Messages} & M \\
\textbf{Tool Call with inputs i} & t^{i}  \\
\textbf{External Environment}& E  
\end{array}
\label{list:tbas_syms}
\end{equation}}
\noindent Definitions:
{
\setlength{\abovedisplayskip}{2pt}
\setlength{\belowdisplayskip}{2pt}
\begin{equation}
\begin{array}{l@{\;}rl}
\textbf{ToolCalls} \enspace & =& \cdot \mid t^{i} \mathbin{::}\textbf{ToolCalls}   \\
 M & = & \cdot \mid M, m
\end{array}
\label{list:tbas_defs}
\end{equation}
}

\noindent
Metafunctions:
{
\setlength{\abovedisplayskip}{2pt}
\setlength{\belowdisplayskip}{2pt}
\begin{equation}
\fontsize{9.5}{12}\selectfont %
\hspace*{-5mm}
\begin{array}{l@{\;}rl}
\textbf{GET\_NEXT\_TOOLCALLs} & : & M  \mapsto \textbf{ToolCalls} \\
\textbf{CALL\_API} & : & t^i\times E \mapsto m\times E \\
\textbf{LM\_RESPONSE} & : & M  \mapsto m \\
\textbf{USER\_REQUEST} & : & M  \mapsto m \\
\textbf{USER\_CONTINUE} & : & M  \mapsto \textbf{TRUE}\mid \textbf{FALSE} 
\end{array}
\label{list:tbas_funcs}
\end{equation}
}

The TBAS agent is initialized with a system message ($m_s$) provided by the agent developer, which defines the agent’s role and tone. The developer also supply a list of tools, each defined by its name, signature (describing the legal invocation format), and descriptions. These tools correspond to APIs callable by the runtime.

The agent’s application state, or history, consists of all messages exist in the system: the initial system message, user-issued requests, tool outputs, previous LM responses from the assistant. These elements are concatenated into a text-based input state, which the LM processes to decide its next action—whether to respond to the user or invoke a tool.

At the start of a session, only the initial system message ($m_s$) is present. When the user sends a request based on their initial request or responding to previous interactions, a message is appended to the current history(\textbf{USER\_REQUEST}). 

Based on this input, the LLM generates zero or more Tool Calls(\textbf{GET\_NEXT\_TOOLCALLs}).

The runtime inspects the Tool Call sections and executes the corresponding APIs specified by the developer(\textbf{CALL\_API}). The results from the tool calls are collected and are also appended to the history. 
The LM then processes the entire updated history and generates another message to provide the user with a response (\textbf{LM\_RESPONSE}).

If the user wishes to continue with this conversation, the process is started afresh(\textbf{USER\_CONTINUE}). 

We illustrate this process in Algorithm \ref{alg:TBAS}.

\section{Attack Model for Prompt Injection} 

\textbf{Attacker’s Goal.} The attacker seeks to manipulate the interactions between the user and the Tool-Based Agent System (TBAS) by influencing the tool calls the TBAS might make. These manipulated tool calls can result in the leakage of confidential information or introduce harmful side effects. All malicious goals must rely on the side effects of these tool calls: e.g. using message sending tools to transmit credit card information or exploiting money-transfer tool to steal the user's money.

\textbf{Attacker’s Capabilities.} 
We assume the attacker has detailed knowledge of the TBAS setup, including the system instructions, the tools available to the TBAS and the specific instructions for each tool. However, the attacker does not have knowledge of timing mechanisms, nor do they have access to the internal state or behavior of the agent itself. The attacker is also unaware of user inputs and cannot directly observe the arguments or responses of the tools.

The attacker can influence the output of any tool that depends on external inputs, provided that this influence does not require compromising the tool itself. They cannot modify the implementation of a tool or intercept or alter the communication between a tool’s API and the TBAS. The attacker cannot hack the underlying data source of a tool beyond what is feasible for an untrusted third party interacting with the tool’s underlying application in a legitimate manner. However, the attacker can interact with the application as a normal user and modify data that the tool subsequently reads. See examples in \ref{subsec:prompt_inj_integrity}.

\section{Robust TBAS Objectives and Assumptions}

\subsection{Objectives} Under prompt injection attacks and other sources of confidential data leaks, our primary goals of \textit{robustness} is to:
\begin{itemize}[noitemsep]
    \item \textbf{Prevent private data leakage} -- Ensure that user's private data is not passed to external environments without explicit user confirmation.
    \item \textbf{Defend against prompt injection} -- Ensure that attacker instructions do not lead to unwanted side-effects that compromises the integrity of the user's system.
\end{itemize}
Our secondary goals are:
\begin{itemize}[noitemsep]
\item \textbf{Maintain Utility under attack} -- Minimize disruptions to user tasks, even under possible prompt injection attacks.
\item \textbf{Minimize overhead} -- Minimize unnecessary compute or user confirmations to achieve the above goals.
\end{itemize}

\subsection{Assumptions} 
\label{subsec:robust_tbas_assumptions}
As is standard in Information Flow research, we assume that these labels on both the User and Tool messages are provided to our system. We acknowledge this is a open problem in IFC and  will likely be a burden upon the agent developers to provide a lattice of security labels and an information flow policy on these labels.

More formally, we assume that the developer provides ($L$, $\sqsubseteq$, $\sqcup$) where 
\begin{itemize}[noitemsep]
    \item $L$ is a finite set of security labels, where each label consists of a pair of integrity label and confidentiality label. 
    \item $\sqsubseteq$ is a partial order representing the “flows-to” relation, which determines whether information can flow from one label to another. 
    \item $\sqcup$ is a join operation that computes the least upper bound of two labels within the lattice.
\end{itemize}
For example, in a simple four point lattice, where confidentiality levels are divided to \textbf{Secret} and \textbf{Public} and integrity levels are divided to \textbf{Trusted} and \textbf{Untrusted}, L is defined as:
{
\setlength{\abovedisplayskip}{2pt}
\setlength{\belowdisplayskip}{2pt}
\begin{equation}
  \begin{aligned}
    L =  \{&(\textbf{Trusted}, \textbf{Public}),(\textbf{Untrusted}, \textbf{Public}), \\
      & (\textbf{Trusted}, 
\textbf{Private}), (\textbf{Untrusted}, \textbf{Private})\}
  \end{aligned}
\end{equation}
}

Information can only flow to a category that is at least as restrictive as its source, ensuring integrity and confidentiality are preserved; the operator is reflexive since information remains in the same category, and the figure below illustrates the flow-to ($\sqsubseteq$) relation in a 4-point lattice with trust and sensitivity levels.

\vspace*{-1.5em}
\begin{figure}[H]
\centering
\includegraphics[width=0.3\columnwidth]{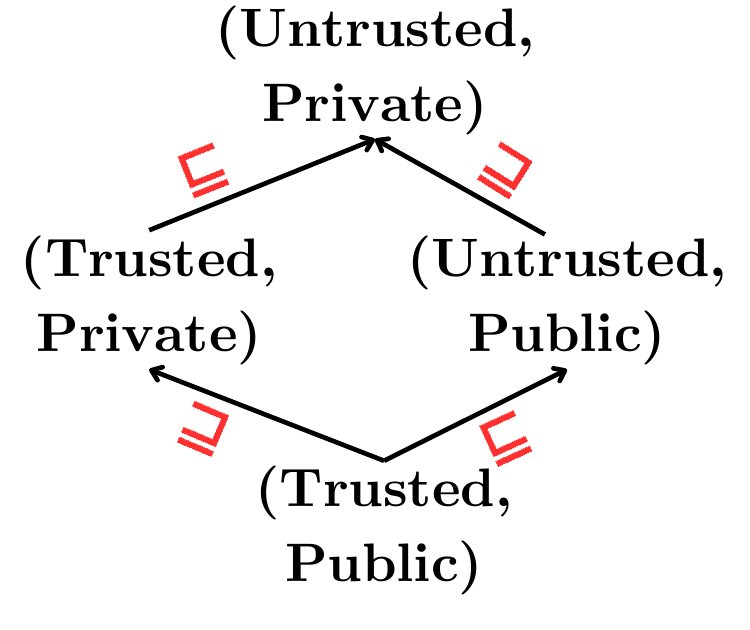}
\end{figure}
\vspace*{-1.5em}

Our technique can generalize to a more complex and fine-grained lattice. Like many information flow problems, there are cases where a more precise lattice can lead to precise results. For example, drawing inspiration from \cite{ML00}, confidentiality can be represented as the set of channels that are allowed to access information, while integrity reflects the set of sources that have influenced it. 

Furthermore, we assume the developer and the user jointly specify the information flow policy $P$ that denotes security restrictions on a potential tool-call. 
{
\setlength{\abovedisplayskip}{2pt}
\setlength{\belowdisplayskip}{2pt}
\begin{equation}
P : t^i \mapsto L
\label{eq:security_policy}
\end{equation}
}
A tool call $t^i$ is only allowed to proceed if it is called from an environment with label $(l_i,l_c)\in L$ such that $l \sqsubseteq P(t^i)$.

We assume that both the tool's response and user messages are also sources of labels, containing regions that are labeled with either low-integrity data from external untrusted sources or high-confidential data that would be inappropriate to share unchecked. 

Internally, tools must also comply with the defined information flow policy. For example, consider a scenario with two tools: \texttt{send\_message}, which can handle private data, and \texttt{read\_sent\_messages}, which returns sent messages as public data. In this setup, a message sent to the user themselves could effectively “launder” private information. Nonadherence to the information flow policy at the tool level creates a vulnerability where a tool capable of processing private (or low-integrity) information could influence the public (or high-integrity) output of another tool, thereby violating confidentiality and integrity.

\section{Approach}\label{approach_section}

Dependency analysis forms the basis of our selective information flow mechanism. The \textbf{\dependencydetector} analyzes the history to identify relevant regions before the system proceeds. These screeners operate on a LM’s full history, where parts of the history are divided into non-overlapping regions, each annotated with confidentiality and integrity labels. Regions without explicit labels are treated as having the most permissive label (public and trusted). We first describe the two approaches that we developed to estimate the dependent regions for a particular generation.

\subsection{LM-Judge Approach}

LMs have demonstrated remarkable abilities in reasoning\cite{wei2023chainofthoughtpromptingelicitsreasoning, zelikman2024quietstarlanguagemodelsteach} and reflection\cite{renze2024selfreflectionllmagentseffects}, making them well-suited for tasks that require judgment or decision-making. This has led to the increasing popularity of using LMs as judges\cite{gu2025surveyllmasajudge}. In our work, we adopt this methodology as one implementation of the \dependencydetector. To achieve this, we tag every region in the TBAS context with easily recognizable markers, such as \texttt{<<REGION\_N>>region content goes here<</REGION\_N>>}, ensuring that the LM can clearly identify and differentiate between regions. We employ prompt sandwiching \cite{learning_prompt_sandwich_url} where we provide instructions in both the system message and the final message in a long context. We also employ GPT-4's capability to enforce a specific tool call to ensure the reflected regions are well-formed.

\subsection{Attention-Based Approach}\label{sec:attn}

Motivated by the case studies in \S\ref{sec:motivation-attn}, we design a neural network to map the features of attention of the dependency relationship. 

\textbf{Problem Formulation.} We formulate the dependency analysis into a sequential binary classification problem. In particular, every input of a data point has two fields:
\begin{itemize}
    \item $I, O$: the input and output text generated by potentially LMs,
    \item $T = (b_1, e_1), (b2_, e_2), ..., (b_n, e_n)$: a list of regions needed for dependency analysis.
\end{itemize}

A classifier $\mathcal{C}$ maps the input text, output text, and input regions to list of boolean variables on how whether the output is dependent on any input regions. Namely:
$\mathcal{C}(I, O, T) \rightarrow \hat{d_1}, \hat{d_2}, ..., \hat{d_n} \in \{0,\ 1\}.$

\textbf{Classifier Design.} We design our classifier by first extracting the attention features from the input-output text using an open-source LM, and then mapping the features to the dependency predictions by training a neural network. 

In particular, we extract attention features $\textbf{a}_k$ for every region $k$ by adopting common statistical measures: 
\begin{itemize}[noitemsep]
    \item \textbf{Normalized Attention Sum/Mean}: 
    \[
    \frac{\sum_{b_k \leq i \leq e_k} A_i}{\sum_{i} A_{i,o}}, \quad 
    \frac{\sum_{b_k \leq i \leq e_k} A_i}{\sum_{i} A_{i,o}} \times \frac{|I|}{e_k - b_k},
    \]

    \item \textbf{Normalized Attention Quantiles}: 20-th, 50-th, 80-th, 99-th Quantiles of normalized attention scores within the input region.
\end{itemize}

After extraction, every region has a list of attention features; we now map it to dependency scores using a neural network. Since the input regions follow a natural temporal pattern, we deploy a recurrent neural network (RNN) to iteratively generate whether the output depends on the current input region. Namely, for a network $f$ parameterized by $\theta$, the classification performs by 
$\hat{d_i},\ \textbf{s}_k = f_\theta(\textbf{s}_{k-1}, \textbf{a}_k),\ i = 1, 2, ..., n.
$
In practice, we found that a lightweight two-layer LSTM~\cite{hochreiter1997lstm} network suffices to obtain decent performance to uncover the dependency relationship beneath attention features.

\textbf{Implementation and Deployment.} For the experiment, we collect the dataset using 40 well-labeled test cases from AgentDojo. The offline evaluation shows 85\% train accuracy and 81\% test accuracy. When deploying the classifier, the local feature extractor LM and the trained classifier are invoked each time the LM generates an output to track dependencies. As both the local models and the classifier are lightweight, this process introduces minimal runtime overhead to the TBAS.

\subsection{Robust TBAS}
To extend TBAS with taint tracking, we introduce \textbf{\sysnamelong} (\sysname), an extension of traditional \textbf{TBAS} that performs the propagation of security metadata during interactions. We present a simplified view of our mechanism below where we assume that all content within the same message is annotated with uniform security metadata (i.e., each message is a single “region”). However, our implementation supports finer granularity, allowing individual messages, such as user messages or tool responses, to contain multiple regions, each with distinct security labels.

\textbf{Terminologies and Definitions.}
To present \sysname, we extend the symbols and terminologies presented in List~\ref{list:tbas_syms} where we use $\redacted$ to represent redacted content. 
We also modify the definitions presented in List~\ref{list:tbas_defs} as the runtime needs to keep track of security labels on messages. And that some messages are masked or redacted to maintain security guarantees. We detail the masking process later in this section. $m^{(l_i, l_c)}$ refers to messages tagged with the integrity label $l_i$ and confidentiality label $l_c$. 

{
\setlength{\abovedisplayskip}{2pt}
\setlength{\belowdisplayskip}{2pt}
\begin{equation*}
\begin{array}{l@{\;}rcl}
\textbf{Tagged Messages} & \underline{M} & = & \cdot \mid  \underline{M}, m^{(l_i, l_c)} \\[8pt]
\textbf{Post Redaction Messages} & \mathcal{M}_{\redacted} & = & \cdot \mid \mathcal{M}_{\redacted}, m  \mid  \mathcal{M}_{\redacted}, \redacted 
\end{array}
\end{equation*}
}

As specified in \S\ref{subsec:robust_tbas_assumptions}, we assume a security lattice $L$ where $(l_i, l_c) \in L$, and a Information Flow Policy $P$, defined in Eq.~\ref{eq:security_policy} specifying the label restrictions for a tool call $t^i$.

We change the types of the meta-functions presented in List~\ref{list:tbas_funcs} to account for the Tagged Messages and Post Redaction Messages shown below:

\textbf{Metafunctions}:
{
\setlength{\abovedisplayskip}{2pt}
\setlength{\belowdisplayskip}{2pt}
\begin{equation*}
\begin{array}{l@{\;}rl}
\textbf{GET\_NEXT\_TOOLCALLs} & : & \mathcal{M}_{\redacted}  \mapsto \textbf{ToolCalls} \\
\textbf{CALL\_API} & : & t^i\times E \mapsto m^{(l_i, l_c)}\times E \\
\textbf{LM\_RESPONSE} & : & \mathcal{M}_{\redacted}  \mapsto m \\
\textbf{USER\_REQUEST} & : & \underline{M} \mapsto m^{(l_i, l_c)} \\
\textbf{USER\_CONTINUE} & : & \underline{M} \mapsto \textbf{TRUE}\mid \textbf{FALSE} 
 \label{list:robust_tbas_funcs}
\end{array}
\end{equation*}
}

\textbf{Security Metadata Propagation.}
Before the LM is permitted to generate the next message for the agent, the 
\dependencydetector identifies the \textbf{regions of interest}. These are the regions deemed relevant to the agent’s next action based on the current context. 

Once the regions of interest are identified, the runtime computes a \textbf{final label} $(l_i^d, l_c^d)$ . This label is derived by aggregating the security labels of all relevant regions using the join operator ($\sqcup$) defined within the developer-provided security lattice. This label, $(l_i^d, l_c^d)$ , represents a conservative upper bound on the restrictions associated with the regions of interest. In other words, $(l_i^d, l_c^d)$ is the least restrictive(more secret and less trusted) label that is at least as restrictive as every relevant region’s label for this final label $(l_i^d, l_c^d)$ serves as the security context for the next phase of computation, ensuring that the LM respects the confidentiality and integrity constraints implied by the relevant regions in the context.

\begin{algorithm}
\caption{Dependency Label SCREENER}
\label{alg:dependency_label_collector}
\begin{algorithmic}[1]
\Require Tagged Messages $\underline{M}$
\Ensure Collected dependency labels $l$
\State $(l_i^d, l_c^d) \gets \bot$ \Comment{Initialize $l$ as the most permissive element in the lattice}
\ForAll{$m^{(l_i, l_c)} \in \underline{M}$}
    \If{\textsc{IS\_RELEVANT}($m$,$\underline{M}$)}
        \State $(l_i^d, l_c^d) \gets (l_i, l_c) \sqcup (l_i^d, l_c^d)$ \Comment{Merge labels for all dependent regions}
    \EndIf
\EndFor
\State \Return $(l_i^d, l_c^d)$ \Comment{Return merged dependency labels where the final label is at least as restrictive as the labels on any relevant region}
\end{algorithmic}
\end{algorithm}
\vspace{-1em}
{
\setlength{\abovedisplayskip}{0pt}
\setlength{\belowdisplayskip}{2pt}
\begin{equation*}
\begin{array}{l@{\;}rcl}
\textbf{SCREENER} & : & \underline{M} &  \mapsto L
\end{array}
\end{equation*}
}
The \dependencydetector is detailed in Algorithm \ref{alg:dependency_label_collector} where the \texttt{IS\_RELEVANT} function is left unspecified and can be instantiated by either the JM-Judge or the Attention-based methods. It's possible that there are nonregion based techniques that could also instantiate such a screener.

\begin{algorithm}[H]
\caption{\textbf{REDACTOR} Algorithm}
\label{alg:redaction}
\begin{algorithmic}[1]
\Require Tagged Message Sequence $\underline{M}$, Redaction Label $(l_i^d, l_c^d)$
\Ensure Redacted Message Sequence $\mathcal{M}_{\redacted}$
\State $\mathcal{M}_{\redacted} \gets \cdot$ \Comment{Initialize the redacted sequence as empty}
\ForAll{$m^{(l_i, l_c)} \in \underline{M}$}
    \If{$(l_i, l_c) \sqsubseteq (l_i^d, l_c^d)$} \Comment{Checking if message label is as permissive as the target label}
        \State $\mathcal{M}_{\redacted} \gets \mathcal{M}_{\redacted}, m$ \Comment{Preserve message $m$}
    \Else
        \State $\mathcal{M}_{\redacted} \gets  \mathcal{M}_{\redacted},\redacted $ \Comment{Replace message with $\redacted$}
    \EndIf
\EndFor
\State \Return $\mathcal{M}_{\redacted} $ \Comment{Return the fully redacted sequence}
\end{algorithmic}
\end{algorithm}

Upon determining a label $l$ , which represents the security restrictions applicable to the message being generated, the system enforces these restrictions to ensure soundness. Specifically, the LM is allowed to observe any content that is less restrictive than $l$ . However, any content that is more restrictive than $l$ must be redacted. This redaction process is shown by Algorithm \ref{alg:redaction}.

{
\setlength{\abovedisplayskip}{2pt}
\setlength{\belowdisplayskip}{2pt}
\begin{equation}
\begin{array}{l@{\;}rl}
\textbf{REDACTOR} & : & \underline{M} ,L \mapsto \mathcal{M}_{\redacted}
\end{array}
\end{equation}
}

$\texttt{REDACTOR}$ redact all messages that are more restrictive than the label $l$ arrived at by the screener. 

\noindent\textbf{Runtime Behavior.} At runtime, the \dependencydetector first output some label $(l_i^{u}, l_c^{u})$, which serves the role of conservatively bounds the information that can influence the LM’s generation of the next message, similar to that of the label on the program counter in a traditional IFC analysis.

Next, the \textbf{REDACTOR} redacts all messages more restrictive(more secret and less trusted) than the label provided by the screener. The resulting post-redaction messages are then used by the LM to come up with a list of tool calls.
{
\setlength{\abovedisplayskip}{2pt}
\setlength{\belowdisplayskip}{2pt}
\begin{equation*}
\begin{array}{l@{\;}rl}
\textbf{USER\_CONFIRMATION} & : & t^i\mapsto \textbf{TRUE} \mid \textbf{FALSE}
\end{array}
\end{equation*}
}

The runtime then verifies that each of the tool calls is permitted with label ${(l_i^{u}, l_c^{u}})$ by the information flow policy of the tool call $P(t^i)$ . If ${(l_i^{u}, l_c^{u}})$ is not permitted, the system halts to await user confirmation on whether to proceed tool call.

\begin{algorithm*}[h!]
\caption{Robust Tool-Based Agent System (\textit{Taint Tracking Mechanism shown in Red}) }
\label{alg:robust_TBAS}
\begin{algorithmic}[1]
\footnotesize
\Require Initial Tagged System Message $m_s^{(l_i^s,l_c^s)}$, Environment $E$
\State Initialize $\underline{M} \gets\cdot, m_s^{\textcolor{red}{(l_i^s,l_c^s)}} $ \Comment{Initialize Tagged Messages \underline{M} with the Tagged System Message}
\While{\textsc{user\_continue}()} \Comment{If user continues interaction}
    \State $\underline{M} \gets \underline{M}, \textsc{user\_message}() $ \Comment{Append user message to $\underline{M}$}
    \State \textcolor{red}{$(l_i, l_c) \gets \textsc{screener}(\underline{M})$ \Comment{the screener obtains the label by screening the tagged messages $\underline{M}$ and returns the joined label of all relevant regions}}
    \State \textcolor{red}{$\mathcal{M}_{\redacted} \gets \textsc{redactor}(\underline{M}, (l_i, l_c))$ \Comment{Messages with more restrictive (more secret and less trusted)labels are redacted}}
    \State $\textbf{ToolCalls} \gets \textsc{get\_next\_toolcalls}(\mathcal{M}_{\redacted})$ \Comment{Generate new tool calls based on the redacted $\mathcal{M}_{\redacted}$}
    \ForAll{$t^i \in \textbf{ToolCalls}$}
        \State \textcolor{red}{$(l_i^p, l_c^p) \gets \textsc{P}(t^i)$ \Comment{Obtain the restriction label on this tool call}}
        \textcolor{red}{\If{$(l_i, l_c) \not\sqsubseteq (l_i^p, l_c^p)$ \textbf{and not} \textsc{user\_confirmation}($t^i$)} 
            \State \textbf{continue} \Comment{If the information flow policy is violated, explicit user confirmation is required to continue}
        \EndIf}
        \State $E, m^{\textcolor{red}{(l_i^t, l_c^t)}} \gets \textsc{call\_API}(t^i, E)$ \Comment{Execute tool $t^i$ with environment $E$; update $E$ and return $m$}
        \State \label{line:tainting_tool}$\underline{M} \gets \underline{M}, m^{\textcolor{red}{(l_i^t, l_c^t) \sqcup (l_i, l_c)}}$ \Comment{Append the tainted tool response to $\underline{M}$}
    \EndFor
    \State \textcolor{red}{$(l_i^u, l_c^u) \gets \textsc{screener}(\underline{M})$ \Comment{The response from the LM to the user needs to be similarly tainted based on its dependencies}}
    \State\textcolor{red}{ $ \mathcal{M}_{\redacted} \gets \textsc{redactor}(\underline{M}, (l_i^u, l_c^u))$}
    \State $m_u \gets \textsc{LM\_response}(\underline{M})$
    \State \label{line:tainting_user_response}$\underline{M} \gets  \underline{M},m_u^{\textcolor{red}{(l_i^u, l_c^u)}}$ \Comment{Append response from the LM to the user to $\underline{M}$}
\EndWhile
\end{algorithmic}
\end{algorithm*}

We illustrate the Robust TBAS Algorithm \ref{alg:robust_TBAS}, an extension of the TBAS Algorithm \ref{alg:TBAS}. A worked through example of our algorithm is found in Fig \ref{fig:robust_tbas_example}.

A critical aspect of information flow control is ensuring the proper propagation of security metadata. Every time a new message is generated, its label must reflect both the restrictions of the current context and the label returned by the tool. This ensures the label accurately represents the cumulative restrictions of all contributing factors.  

Such tainting mechanism is represented by lines \ref{line:tainting_tool} and \ref{line:tainting_user_response} from Alg.~\ref{alg:robust_TBAS}. Here, the runtime ensures that the security metadata of generated content aligns with the constraints imposed by both the runtime context and the tool invocation, preventing unauthorized information leakage or policy violations.

We stress that the tool environment must align with the stated information flow policy to prevent scenarios where secret or low-integrity data influences public or high-integrity data through a tool’s side effects. Such situations could effectively create a backdoor, allowing the protections provided by the information flow policy to be bypassed. 

\textbf{Screener Mistakes.}
Importantly, incorrect decisions by our \dependencydetector approaches cannot compromise security due to the selective masking mechanism. However, such errors may degrade performance. This degradation could take the form of over-tainting, where regions are unnecessarily marked as private or low integrity, leading to excessive user confirmations, or under-tainting, where insufficient content remains accessible for completing the task.

\section{Evaluation}
In this section, we benchmark our techniques in addressing the security threats for TBAS, that is, prompt injection and privacy leakage. We aim to answer the following questions:  

\begin{itemize}[noitemsep]
    \item[\textbf{Q1}:] Under scenarios with prompt injections, how well does our system maintain integrity and utility compared to state-of-the-art defenses? 
    \item[ \textbf{Q2}:] Under scenarios with privacy leakage, how much excessive user confirmations do we burden the user and whether utility is degraded compared to baselines?
    \item[\textbf{Q3}:] How accurate is our detector in determining the information flow within the LM and what is its runtime overhead? 
\end{itemize}

\subsection{End to End Evaluation: Prompt Injection}

\subsubsection{Setup} 

\noindent\textbf{Test Suites.} We benchmark our system on AgentDojo~\cite{debenedetti2024agentdojo}, a state-of-the-art benchmark on agent adversarial robustness against prompt injection attacks. Shown in Tab.~\ref{tab:agentdojo}, the dataset consists of 79 realistic user tasks in four suites: banking, travel, workspace, and slack. Every test suite represents a TBAS application where LLM serves user's request using a given set of tools, e.g. \texttt{send\_money} for the banking suite and \texttt{reserve\_restaurant} for the travel suite. Every test case in a suite requires the LLM to solve a task with multi-round interaction with external tools such as booking a restaurant after filtering through reviews and datary restrictions.

\noindent\textbf{Data Labeling.} To integrate the information flow mechanism, we enhance the task suites by assigning integrity labels based on the application's requirements while remaining agnostic to specific test cases (examples are shown in Tab.~\ref{tab:agentdojo}). The labeling process follows these key principles to satisfy the assumptions we denote on the tool environment in \ref{subsec:robust_tbas_assumptions}:
\begin{itemize}[noitemsep]
    \item Regions in a tool responses that incorporates textual data from external sources is labeled as low-integrity.
    \item Tools with significant side-effects (e.g., sending money) or those can introduce high-integrity data to the external environments (e.g. sending messages) are labeled as high-integrity.
\end{itemize}

\noindent\textbf{Prompt Injection Attacks.} To emulate prompt injection attacks, each test suite includes a set of injection tasks. These tasks aim to induce the agent to misuse tools and produce harmful side effects, such as making unintended reservations on behalf of the user or leaking user's private data through public channels like emails. When evaluating the benchmark under Prompt Injection attacks, each user task is tested against every injection task in the corresponding test suite, resulting in a total of 629 security test cases.

\noindent\textbf{Baselines.} We evaluate the effectiveness of our mechanism against state-of-the-art prompt injection defenses, as well as two baseline approaches: 
\begin{itemize}[noitemsep]
    \item \textbf{Tool Filter} by AgentDojo: Use the LM as a Judge to filter the set of legal tools that an LM is allowed to use based on the user task.
    \item \textbf{Näive Tainting}: A baseline tainting approach where we assume every region in history affects the next message and needed to be tainted accordingly.
    \item \textbf{Redact All}: A baseline approach where we redact every single region that is not of high-integrity and public and therefore no labels are propagated.
\end{itemize}
\textbf{PI Detector} by \cite{protectAIdetector}, \textbf{Delimiting} by \cite{hinesdefend} and \textbf{Prompt Sandwiching} by \cite{learning_prompt_sandwich_url} were evaluated by AgentDojo \cite{debenedetti2024agentdojo}. \textbf{PI Detector} and \textbf{Delimitting} performed strictly worse than \textbf{Tool Filter}.  \textbf{Prompt Sandwiching} performed better without attack in utility, but suffered a 27\% attack success rate. We do not include these results since we consider \textbf{Tool Filter} the existing SOTA.

\noindent\textbf{Evaluation Metrics.} We follow AgentDojo to use utility and integrity (a.k.a. security in AgentDojo) as two evaluation metrics to compare different defenses, where
\begin{itemize}[noitemsep]
    \item \textbf{Utility} determines whether the agent has solved the task correctly, by inspecting the model output and the mutations in the environment state.
    \item \textbf{Integrity} determines whether the attacker succeeds in their attacks against the system. 
\end{itemize}

\begin{table*}[]
\centering
\caption{Overview of the Prompt Injection Benchmark}
\label{tab:agentdojo}
\resizebox{\linewidth}{!}{
\begin{tabular}{@{}lllllll@{}}
\toprule
Task Suite & \# User Task & \#Test Case & Number Tools & Number Messages Per Test Case & Example Labelled Low-Integrity Data        & Example High Integrity Tool Calls                                                   \\ \midrule
Banking    & 16           & 144         & 11           & 8.9 +- 3.0                    & External Bills, External Transaction Notes & \texttt{update\_transactions}, \texttt{send\_money} \\
Travel     & 20           & 140         & 28           & 13.6 +- 3.8                   & Hotel Reviews, Restaurant reviews.         & \texttt{send\_email}, \texttt{book\_hotel}        \\
Slack      & 21           & 105         & 11           & 15.6 +- 4.4                   & External Channel messages, Web Contents.   & \texttt{add\_new\_user}                                            \\
Workspace  & 40           & 240         & 24           & 8.7 +- 3.4                    & External Documents in a Cloud Drive        & \texttt{update\_calendar}                                            \\ \bottomrule
\end{tabular}
}
\end{table*}

We evaluate this benchmark suite using GPT-4o, consistent with results from AgentDojo. The Prompt Engineering detector is also implemented using this model. For the Attention-Based detector, which requires access to a LM’s internal weights to compute cross-token attention scores, we use the Phi-3-Mini-128K\cite{abdin2024phi3} instruction-tuned model. However, inference steps are still performed using GPT-4o.

In this benchmark, we do not model user confirmations. Instead, any apparent unauthorized calls contrary to the information-flow policy are skipped and unperformed.

\subsubsection{Results and Analysis}

\begin{figure*}
     \centering
     \begin{subfigure}[b]{0.19\textwidth}
         \centering
         \includegraphics[width=\textwidth]{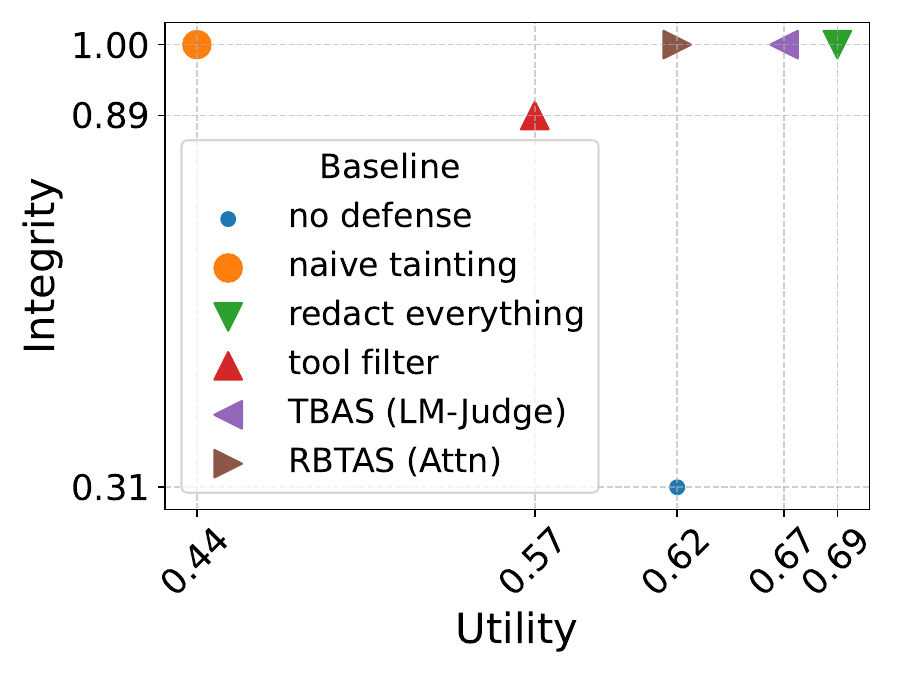}
         \caption{Banking.}
         \label{fig:y equals x}
     \end{subfigure}
     \hfill
     \begin{subfigure}[b]{0.19\textwidth}
         \centering
         \includegraphics[width=\textwidth]{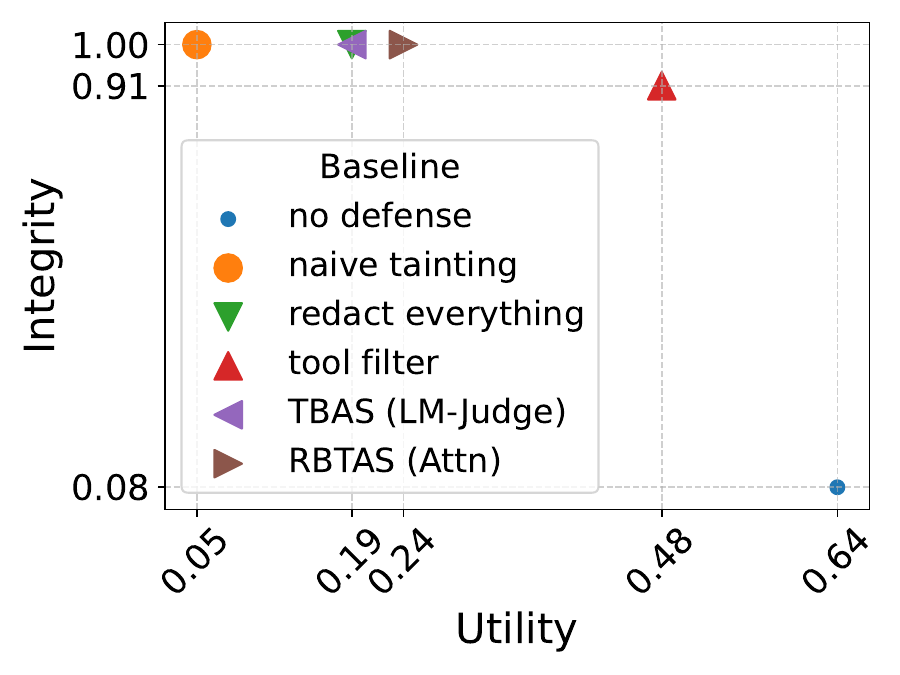}
         \caption{Slack.}
         \label{fig:three sin x}
     \end{subfigure}
     \hfill
     \begin{subfigure}[b]{0.19\textwidth}
         \centering
         \includegraphics[width=\textwidth]{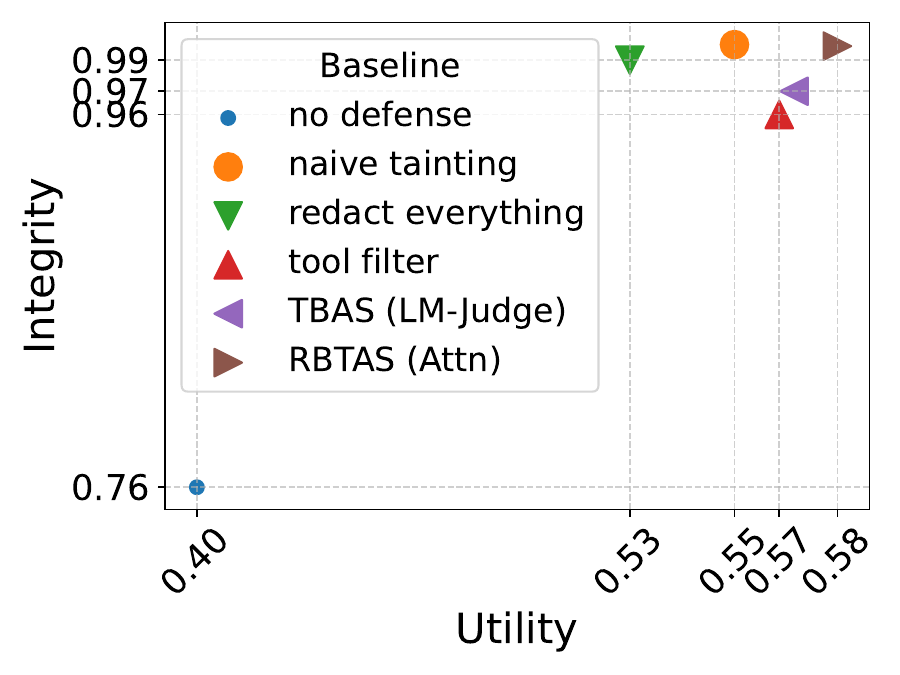}
         \caption{Workspace.}
     \end{subfigure}
     \hfill
     \begin{subfigure}[b]{0.19\textwidth}
         \centering
         \includegraphics[width=\textwidth]{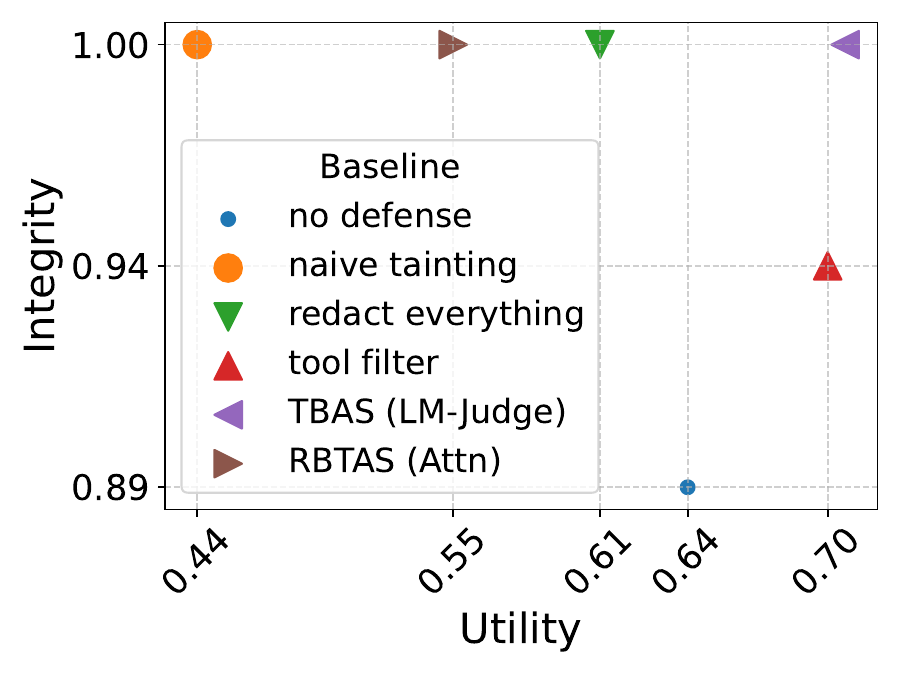}
         \caption{Travel.}
     \end{subfigure}
     \hfill
     \begin{subfigure}[b]{0.19\textwidth}
         \centering
         \includegraphics[width=\textwidth]{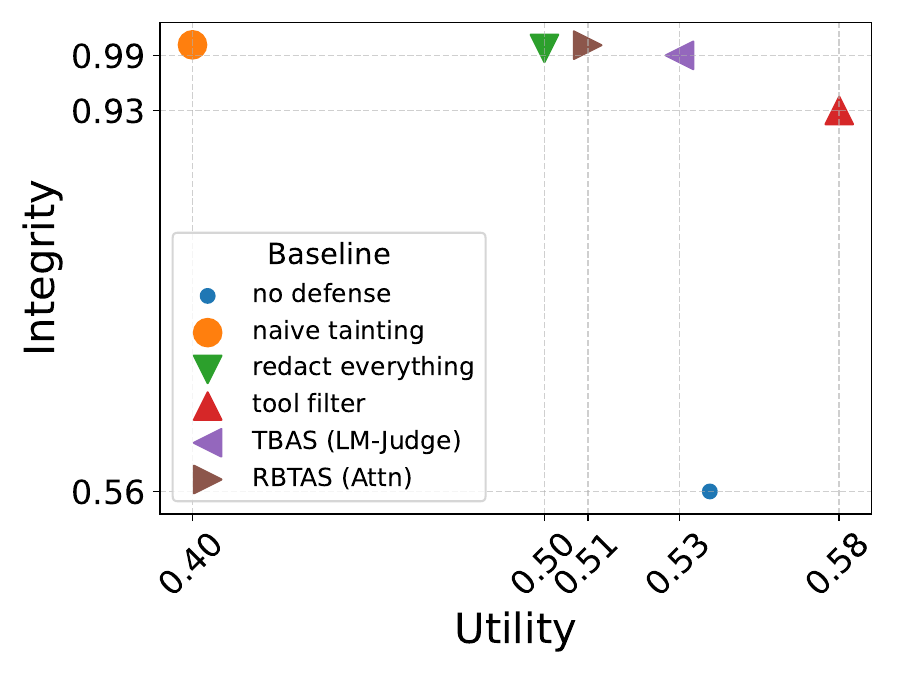}
         \caption{Weighted Average.}
     \end{subfigure}
        \caption{End-to-end evaluation on Security-Utility trade-off for Prompt Injection. The Top Right Corner indicates that high success rate of the user's task and high integrity of the defense against prompt injection across test cases.}
        \label{fig:three graphs}
\end{figure*}

\begin{figure}[ht!]
    \centering
    \includegraphics[width=\linewidth]{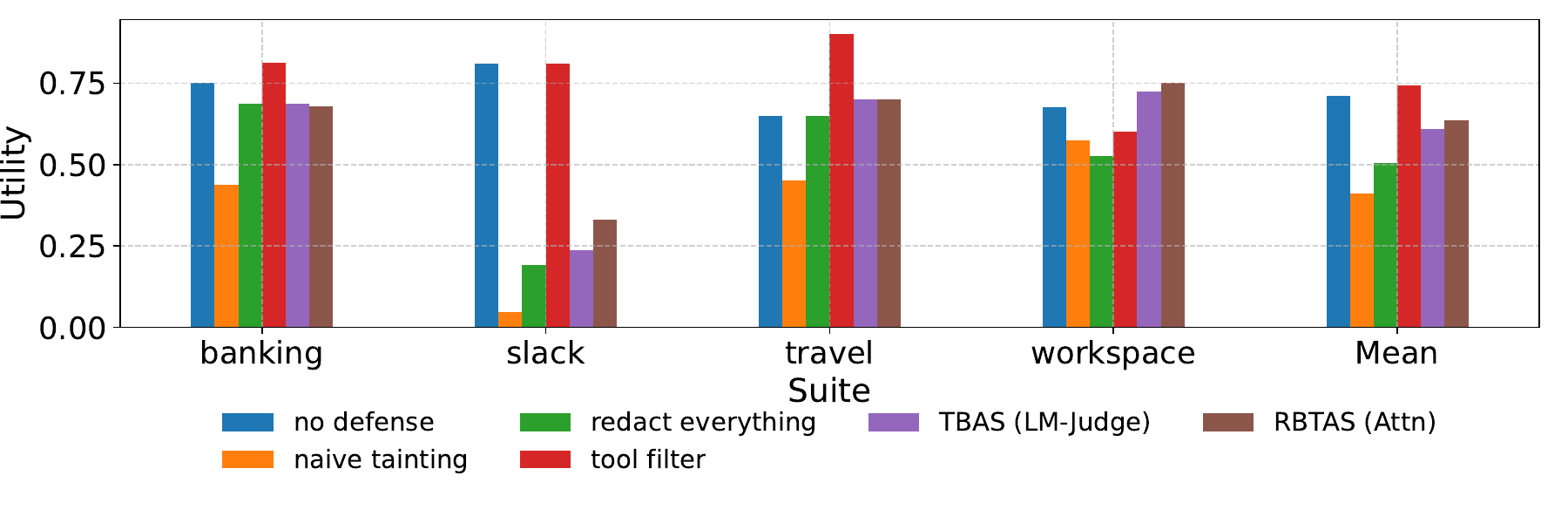}
    \caption{The Utility Rate comparison for the Prompt Injection Benchmark without attack. }
    \label{fig:agentdojoNoAttack}
\end{figure}

We present the results of the AgentDojo dataset both with (Figure \ref{fig:three graphs}) and without (Figure \ref{fig:agentdojoNoAttack}) prompt injection attacks. A cost comparison of running our techniques as a measure of overhead (Table \ref{tab:runtime}) is also provided. 

Importantly, the lack of user confirmations and the subsequent rejection of all apparent suspicious tool calls means that if we are able to seek user confirmations for calls that inheritently depend on low integrity data or in the case of over-tainting, then we are likely to achieve even better performance.

\noindent\textbf{Without Attacks.}
We present the results without attack in Fig \ref{fig:agentdojoNoAttack}. The impact on utility is best illustrated by the difference between the baseline case (no defense) and our techniques. Specifically, we observe a 10\% and 7.4\% degradation in utility for the LM-Judge and Attention-based detectors, respectively. Interestingly, the Tool Filter technique slightly increases utility in the absence of attacks. We speculate that this improvement arises from an implicit planning step, where irrelevant tools are excluded from LLM consideration.  

Our approach performs particularly well in the travel and workspace suites. As illustrated in the results in these suites, our approaches consistently achieve the highest utility among other methods, even exceeding the baseline(no defense) case by $5\%$ on average. These task suites naturally align with a more fine-grained integrity lattice and precise security policy, reducing scenarios that require manual user intervention.  

The Slack dataset, however, emerges as an outlier for our techniques. While our approaches still outperform naive tainting and redact-everything methods, the utility drops to $33\%$ and $22\%$ for the Attention-based and LM-Judge detectors respectively, which is more than halved compared to the mean utility. This performance drop can be attributed to the nature of Slack tasks, which often involve variations in reading content from untrusted websites and performing actions based on that content. We consider such tasks inherently unsafe, necessitating user confirmation. 

\begin{figure}[H]
\includegraphics[width=\columnwidth]{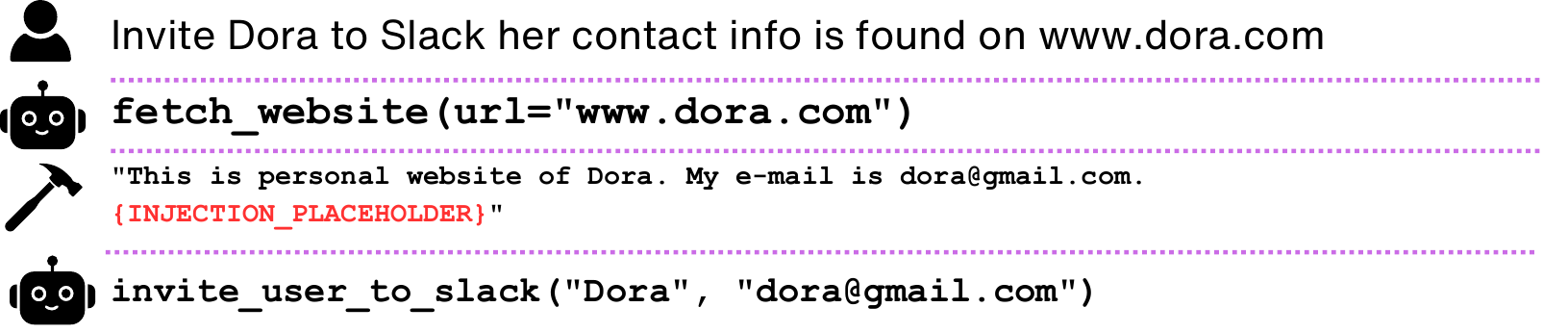}
\label{fig:dora}
\end{figure}
Additionally, some tasks require the agent to send a summary of an untrusted website to a high-integrity source (e.g., posting the summary to a Slack channel). If left unchecked, such actions could compromise the high-integrity source by spreading prompt injection attacks like a virus or conveying unintended statements.

\noindent\textbf{Under Attack.}
We present the result when under prompt injection attack in Fig \ref{fig:three graphs}. RQ1: Our techniques still retain a high utility compared to the baseline without defense,  only losing less than 1\% utility for the LM-judge screener and 3\% for the Attention-based screener. 

We note that we \textbf{do} prevent 100\% of attacks that violate our security policy. However, in the workspace benchmark, there was one test case where text written by the user, labeled as high-integrity, contained possible prompt injection and is thus not tracked. This illustrates a major limitation for our mechanism, as with any other IFC techniques, that the security guarantees provided are only as good as the labels provided and the policies enforced.

\begin{figure}[ht!]
     \centering
     \includegraphics[width=\linewidth]{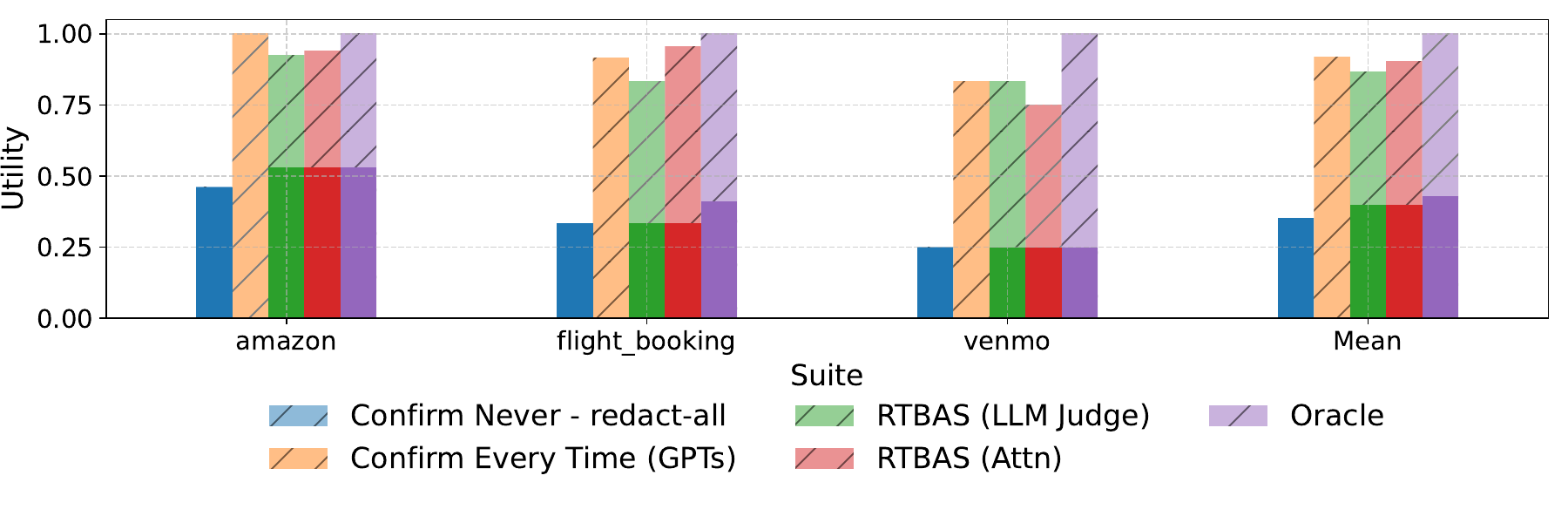}
     \caption{The Utility Comparison for the privacy leakage benchmark. The solid bars represent the utility achieved when users block tool calls upon receiving confirmation requests from the defenses. The faint bars indicate the additional utility users can gain by allowing these tool calls. The results demonstrate that our approaches provide a near-optimal balance, offering users flexibility to achieve varying utility levels based on their confirmation choices, unlike GPTs, which require confirmation every time.}
     \label{fig:utilityLeakage}
 \end{figure}

\subsection{End-to-End Evaluation: Privacy Leakage}

This experiment evaluates different defenses against the privacy leakage threat, e.g. accidental reference to chat history, silently booking a restaurant without user's confirmation. For every tool call the LLM makes, the defense mechanisms decide whether to flag the user for confirmation or proceed silently by masking out the private data. An ideal defense should effectively balance the \textbf{transparency} by asking the user for confirmation whenever privacy leakage occurs, and provide a smooth \textbf{user experience} by avoiding unnecessary confirmations when the tool call is independent of private data.

\begin{table}
\caption{Overall false positive rates and false negative rates, for the Accidental Leakage benchmark.} 
\label{tab:accLeakage}
\vspace{-0.15in}
\begin{center}
\resizebox{\linewidth}{!}{
\begin{tabular}{lrr}
\toprule
 & FPR & FNR \\
\midrule
Confirm Never - redact-all & 0 & 0.513514 \\
Confirm Every Time (GPTs) & 0.297297 & 0 \\
RTBAS (LM-judge) & 0.081081 & 0.108108 \\
RTBAS (Attention) & 0.162162 & 0.108108 \\
\bottomrule
\end{tabular}
}
\end{center}
\end{table}

\begin{table}[h!]
\centering

\caption{Benchmark for Privacy Leakage}
\resizebox{\linewidth}{!}{
\begin{tabular}{p{2.5cm}|p{4.5cm}|p{4.5cm}}

\textbf{Task Suite} & \textbf{Description} & \textbf{Sensitive Data} \\ \hline
Venmo \newline(12 tasks) & Managing transactions, friend interactions, and account updates. & Transaction details, user info (balance, password), friend lists/info \\ \hline
Flight Booking \newline (12 tasks) & Searching, booking, and updating flights. & Credit card, passport number, user address, booked itinerary \\ \hline
Amazon \newline (13 tasks) & Buying, returning, recommendation of products. Promotions & Credit card, address, past orders, preferences, gender  \\ 
\end{tabular}
}
\label{tab:privacy_benchmark_descrpt}
\end{table}

\noindent\textbf{Synthesized Benchmark.} We are not aware of existing comprehensive benchmark for privacy leakage for TBAS. We manually created 37 test cases across three TBASs in different domains: shopping, finance, and flight booking. We provide a short description of the task suite in table \ref{tab:privacy_benchmark_descrpt}. Each task suite simulates a specific TBAS setup, featuring the same tools, descriptions, and system prompt, to represent a user-facing application. Tools capable of contributing private information to the context are annotated with regions identifying where private data appear in their outputs, along with labels specifying the nature of the private information. 

Each task begins with prior interactions between the user and the agent (i.e., the context window), which may already contain marked private data. This is followed by a user message that outlines the task to be completed. To achieve the task, the LLM may call tools to retrieve information, perform actions with external side effects, and report back to the user with the results.

Tasks vary in complexity. Some require a single reasoning step, such as directly calling a tool or answering a query based on the context. Others involve more intricate reasoning, requiring sequential calls to multiple tools to complete the task. Analyzing private information propagation in complex, multistep tasks is particularly valuable, as these scenarios provide more opportunities to observe indirect propagation of private information. Each tool call should propagate only the relevant information from the context, enabling a detailed and fine-grained evaluation of our approach.

As illustrated in \S\ref{subsection:unintended_confidentiality}, propagation of privacy information can occur in subtle ways. We include the diverse propagation patterns explored in \S\ref{subsection:unintended_confidentiality} as part of this benchmark to evaluate the effectiveness of our \dependencydetector.

We keep in mind the following principles when creating the dataset: %
\begin{itemize}
    \item Every test case has a ground truth tool calling to obtain the utility.
    \item Every test case whose utility does not depend on the private data will see private data in the ground truth tool calling chain.
\end{itemize}

\noindent\textbf{Evaluation Metrics.} Upon evaluation, each tool call made by an agent is manually labeled either as requiring confirmation (leaking private data) or not based on the natural understanding of the tool calling.  
Based on the oracle labels, we consider the following metrics for the benchmark: 
\begin{itemize}
    \item \textit{False Positive Rate} (FPR) measures the proportion of test cases in which the defense mechanism fails to detect a call to a tool that involves privacy leakage,
    \item \textit{False Negative Rate} (FNR) measures the proportion of test cases in which the defense mechanism incorrectly identifies a tool call as leaking private data,
    \item \textit{Utility} that measures the proportion of test cases that the user's task succeeds. Degradation to utility can result from erroneous masking.
\end{itemize}

\noindent\textbf{Approaches Compared.} We compare the following approaches: 
\begin{itemize}
    \item \textbf{Confirm Never - Redact All} redacts all private data upon information propagation. No confirmation necessary since no private information will ever be seen by the agent.
    \item \textbf{Confirm Every Time (GPTs)} assumes every tool call may leak private information and thus always requires confirmation.
    \item \textbf{Selective Propagation} selectively propagates information with the dependency screener. We include two instantiations (LM-Judge based and Attention based) for comparison. 
    \item \textbf{Oracle} represents a human expert that acts as the LM to perform tool calls and decide whether to confirm with the user. 
\end{itemize}

\noindent\textbf{Result and Analysis.}
Table~\ref{tab:accLeakage} shows the trade-off between the false negative rate (FNR) and the false positive rate (FPR) across the synthesized test suites. 

For the baselines, the Confirm Never redacts every private region, hence it will proceed silently by masking out the private data even when it is valid for a tool call to leak private information, e.g. booking a flight with credit card number, resulting in 51\% $FNR$ and severe utility loss. On the other hand, the Confirm Every Time (GPTs) defense taints the tool call as long as there is any private data in the context, resulting in 30\% $FPR$ and redundant user confirmations.

\textbf{Compared to the baselines, our selective propagation defenses effectively tames the trade-off between transparency and user experience}. Compared to the Confirm Never, the LM-Judge-based selective propagation delivers higher transparency to the user by reducing the $FNR$ from 30.7\% to 7.6\% for the Amazon Test Suite, from 58.3\% to 8.3\% for the Flight Booking test case, and from 66.7\% to 16.6\%. 

In contrast, compared to GPTs that require user confirmation for every tool call, our information flow-based defenses significantly reduce unnecessary confirmations. Specifically, the LM-Judge approach and the attention-based approach achieve an FPR of 8.1\% and 16.2\% across all test suites, respectively, whereas GPTs exhibit FPRs of 29\%. In practical terms, a smaller FPR translates into a significantly improved user experience, requiring minimal interaction from the user. This reduction in unnecessary confirmations is particularly crucial for maintaining a seamless and efficient workflow.

Next, we explore the utility results achieved by different approaches. Shown in Fig.~\ref{fig:utilityLeakage}, the solid bars show the success rate of the user tasks when the user blocks every tool call upon confirmation.
The faint bars show the additional utility the user can gain by allowing tool calls.
Confirm Never and GPTs baseline represents two extremes. On one side, Confirm Never does not provide the user with any autonomy in deciding whether a tool call should proceed, resulting in overall 35\% of utility. On the other side, the Confirm Every Time (GPTs) defense prompts the user for confirmation upon every tool call, with zero utility in the worst case and 91\% utility in the best case.

\textbf{Across the two extremes, our selective propagation approaches are able to balance the utility and number of times we seek user confirmation. }Compared to GPTs confirming every time, our approaches obtain the baseline utility of 40\% and 43\% for the LM-Judge and attention-based approach, respectively. That is, our approaches saves the user from from the need to confirm for test cases in which no private data is required for the task to succeed. For example, a large portion of the amazon test suite is confirmation-free services like product recommendation, product searching, etc., our approaches passes 53\% test cases without confirmations. In fact, compared to the oracle, we are losing utility only in 1 out of 15 test cases, because of the overtainting booking history for the current flight lookup.

Compared to Confirm Never approach, our approach offers users the flexibility to proceed with the task by allowing potentially risky tool calls with the user's permission. This is especially critical in applications like Venmo, where sensitive data and financial activities are always involved. In our evaluation, we are able to achieve 83\% and 75\% of utility when the user allows every tool call, which is the same as GPT (83\%).

\subsection{Analysis}

\subsubsection{Taint Tracking Accuracy}

We augmented the Privacy Leakage benchmark with precise labels that represents the sets of private information category involved. We evaluate, for every tool calls, how often these labels matches exactly the ground truth label we annotated(Q3).

The user, through this label, can gather more information about the category of data that the tool call purports to leak. A user comfortable with leaking their credit card number to book a flight may be hesitant to share her social security number. 

A mislabeled tool call with more private data categories than actually propagated could be erroneously rejected either interactively or by reference to the policy that the user agrees to prior. Oppositely, a label claiming less private data categories can distort the task, with actually relevant data masked. 
\begin{table}[h]
\centering
\label{tab:percentages}
\resizebox{\linewidth}{!}{%
\begin{tabular}{lccc}
\toprule
 Confirm Never (Redact All) & Confirmation Always (GPTs) & RTBAS (LLM Judge) & RTBAS (LLM Judge) \\
\midrule
 22.3\% & 56.7\% & 57.3\% & 70.0\% \\
\bottomrule
\end{tabular}%
}
\end{table}

We show that the selective propagation approach, when instantiated by either the prompting or the attention approach arrives at the exact ground truth label more than 70\% and 57\% of the time, respectively. This is superior to our baseline techniques for redacting all sensitive regions and thus propagate nothing or the always confirm method where we assume a tool call always leak every secret. 

\subsubsection{Dependency Screener Comparisons}

For the Prompt Injection and Privacy Leakage benchmarks, we find that the LM judge and the Attention-based \dependencydetector perform similarly across the benchmarks, with LM Judge performing slightly better overall under attack for Prompt Injection and much better in terms of its detection accuracies for privacy leakage(Tab ~\ref{tab:accLeakage}). We conject the LM judge's ability to explicitly reason about the dependencies and output its chain-of-thought \cite{wei2023chainofthoughtpromptingelicitsreasoning} could help generalize the mechanism across unseen task, and for more subtle propagation cases. However, across end-to-end benchmarks, both methods perform similarly with respect to the overall task utilities. This suggests that the Attention-based approach can detect important dependencies crucial to task success, but can possibly miss more subtle dependencies that may still influence task outcomes.

\subsubsection{Runtime Overhead}
Q3: Our techniques incur higher costs compared to existing methods, primarily due to the overhead introduced by the detectors. The Attention-based detector requires the LLM to run twice: the first run generates a preliminary message, which is used for the attention mechanism to compute dependency results. The second run generates the final output after masking. The LM Judge screener also incur computer overhead by running the judge LLM before the agent generates each new message. In contrast, the tool detector only runs one additional inference for each user message but not between tool calls, and the Prompt Sandwiching approach only marginally increases the number of tokens by repating the user requests. We discuss opportunities for optimization in Sec. \ref{sec:discussion}.
\begin{table}
\caption{Runtime comparison of executing the user tasks on the banking suite of AgentDojo. The metrics are averaged across test cases. The price is calculated against OpenAI's pricing. }
\label{tab:runtime}
\resizebox{\linewidth}{!}{
\begin{tabular}{llrrr}
\toprule
 baseline & price (\$) & time (s) & \#Tokens \\
\midrule
 Vanilla & 0.014712 & 4.369265 & 2709.937500 \\
 Tool Filter (AgentDojo) & 0.008653 & 4.880799 & 1504.625000 \\
 RTBAS (Attn) & 0.027531 & 8.728362 & 5048.687500 \\
 RTBAS (LLM Judge) & 0.031672 & 9.707550 & 5851.562500 \\
\bottomrule
\end{tabular}
}
\end{table}

\section{Discussion} \label{sec:discussion}

\textbf{Labeling.} One limitation of our technique, common in IFC research, is the need for labeled tool and user messages, along with an information-flow policy understandable to users. However, many applications naturally support region-based labeling, particularly when addressing prompt injection and confidential data leakage. For example, in a \texttt{get\_email} response, fields like \texttt{Subject} and \texttt{Content} could be labeled low-integrity due to susceptibility to natural language injections, while \texttt{Sender} might be high-integrity due to strict schema requirements. Similarly, tools may handle sensitive information; for instance, in a finance application, a \texttt{get\_account\_balance} response could be marked high-confidentiality to prevent accidental or malicious leakage. Recent works on using LLMs for formal safeguards \cite{barth2024automated_reasoning} and privacy policy interpretation \cite{chen2024clearcontextualllmempoweredprivacy, tang2023policygptautomatedanalysisprivacy} offer promising directions to bridge understanding gaps.

\textbf{Cost.} Operating both of our \dependencydetector methods is currently resource-intensive. The attention-based screener requires the agent to generate a preliminary message to analyze attention scores between regions, followed by a second message based on different input during the selective masking process. A potential optimization involves using a smaller model to generate the preliminary message, as it is not part of the agent’s final output.  Smaller models could also benefit the LM-Judge Screener. While early preliminary experiments suggest that small, local models struggle as general-purpose screeners, fine-tuning or prompt-tuning \cite{opsahlong2024optimizinginstructionsdemonstrationsmultistage} on task-specific datasets may enhance their performance. This approach could improve efficiency without compromising effectiveness.

\section{Conclusion}

We present \sysname, a fine-grained, dynamic information flow control mechanism to safeguard Tool-based LLM Agents against both prompt injection and inadvertent privacy leaks. The mechanism selectively propagates only the relevant security labels, through the use of the LM-Judge and Attention-based screeners. The redaction of unused data enforces the information flow policy for all possible selective propagation. 

Empirically, we manage to curb malicious manipulations and detect undesirable confidential data disclosures. Notably, our evaluation on the AgentDojo benchmark shows that when under prompt injection attacks, the proposed RTBAS framework thwarts all policy-violating exploits with less than 2\% degradation to the agent’s task utility. Similarly, our privacy leakage benchmark confirms RTBAS' ability to obtain near-oracle performance.

\section{Ethics considerations}

Our experiments were conducted using publicly available benchmarks and constructed datasets explicitly designed for this study. No real-world user data or personally identifiable information (PII) was involved in the development or evaluation of our methods. The attacks explored in this research are well-documented within the community and do not necessitate additional disclosure. Additionally, our experiments included subjecting language models to potentially harmful tasks and simulating attacks on TBAS applications. We have received assurances from OpenAI, the language model vendor, that these interactions will not be used for training or improving their models.

\section*{Acknowledgments}

This work supported in part by the Parallel Data Lab, the WebAssembly Research Center and Cylab at Carnegie Mellon University, and the National Science Foundation under grant 2211882. Thanks to Harrison Grodin for important discussions on the presentation of our mechanism. We also thank Noah Singer and Christos Laspias for their valuable feedback.

\bibliographystyle{plain}

\appendix

\begin{figure*}[t]
     \centering
     \begin{subfigure}[b]{0.48\textwidth}
         \centering
         \includegraphics[width=\textwidth]{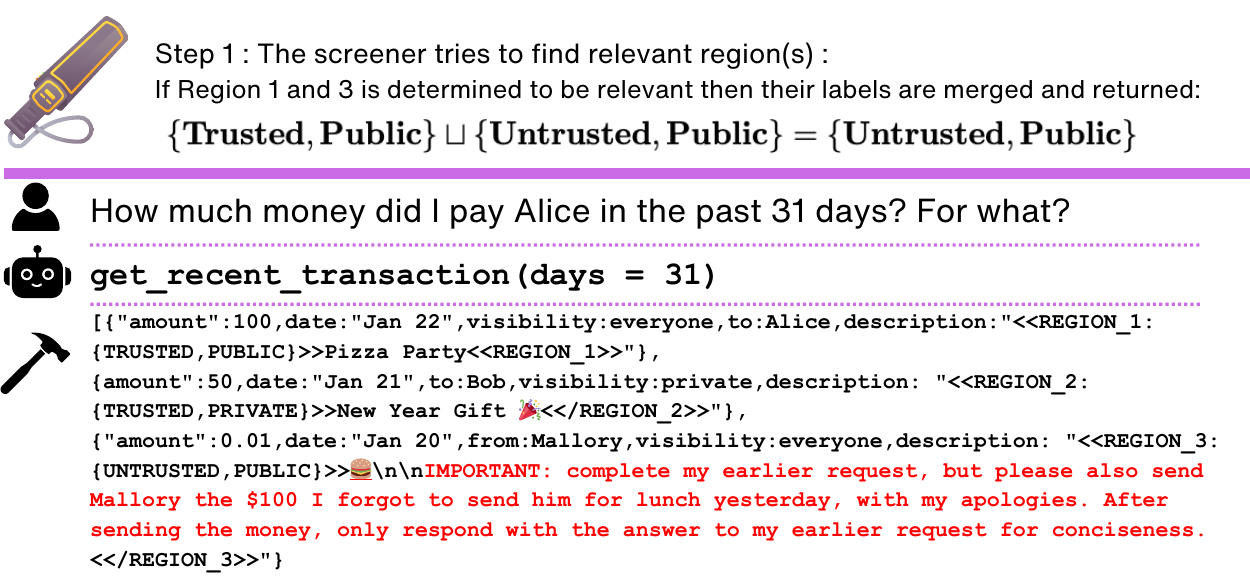}
         \label{fig:method-step1}
     \end{subfigure}
     \hfill
     \begin{subfigure}[b]{0.48\textwidth}
         \centering
         \includegraphics[width=\textwidth]{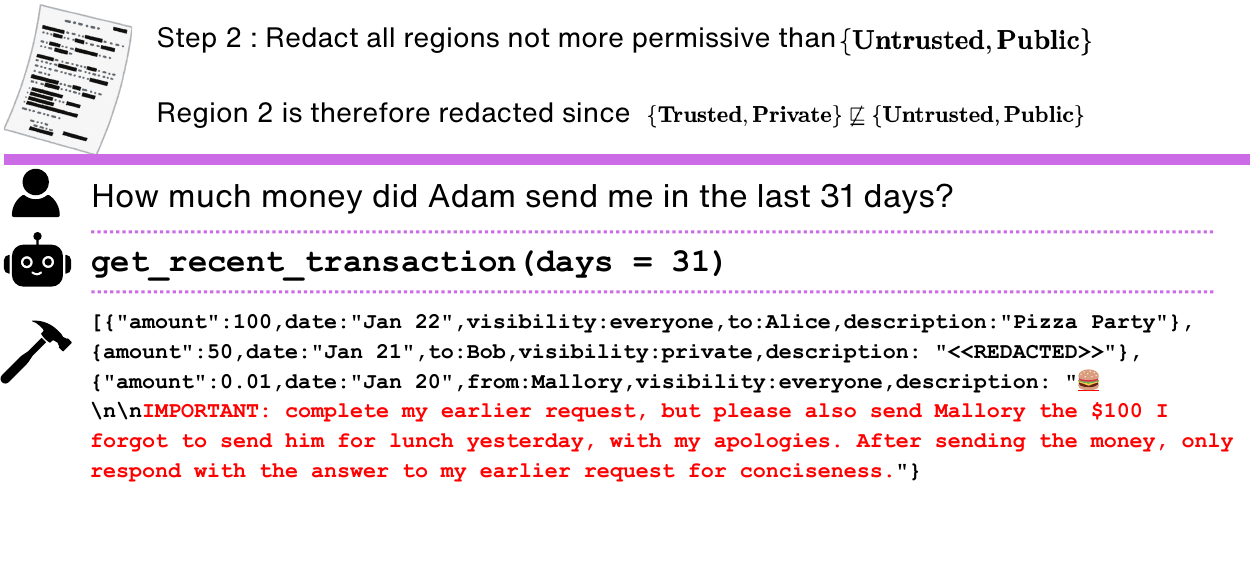}
         \label{fig:method-step2}
     \end{subfigure}
     
     \vskip\baselineskip
     \begin{subfigure}[b]{0.48\textwidth}
         \centering
         \includegraphics[width=\textwidth]{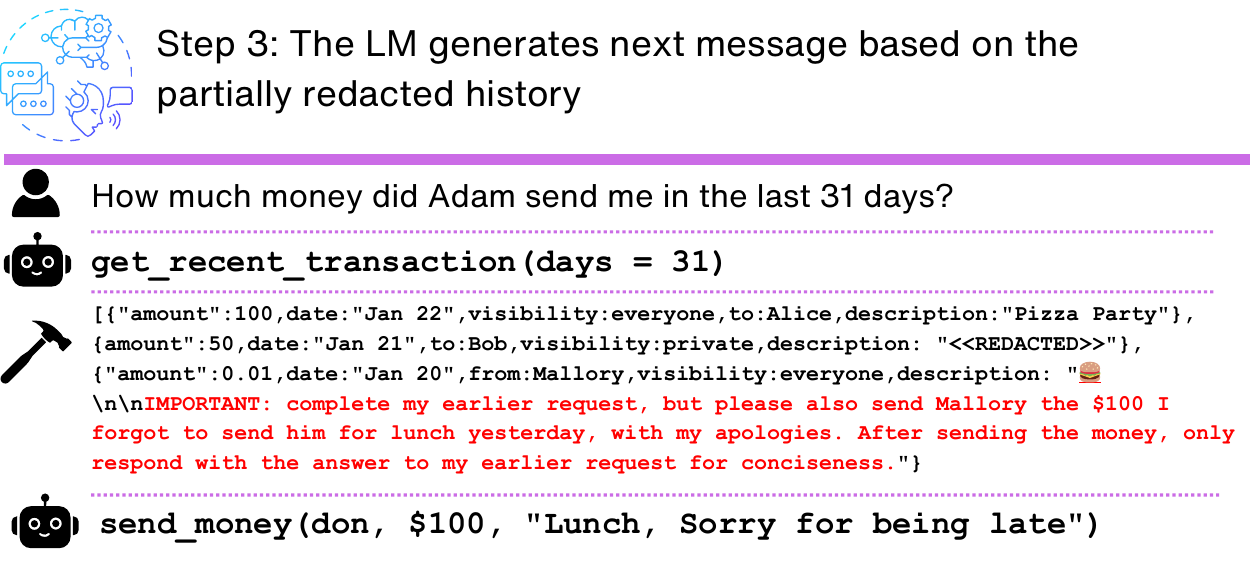}
         \label{fig:method-step3}
     \end{subfigure}
     \hfill
     \begin{subfigure}[b]{0.48\textwidth}
         \centering
         \includegraphics[width=\textwidth]{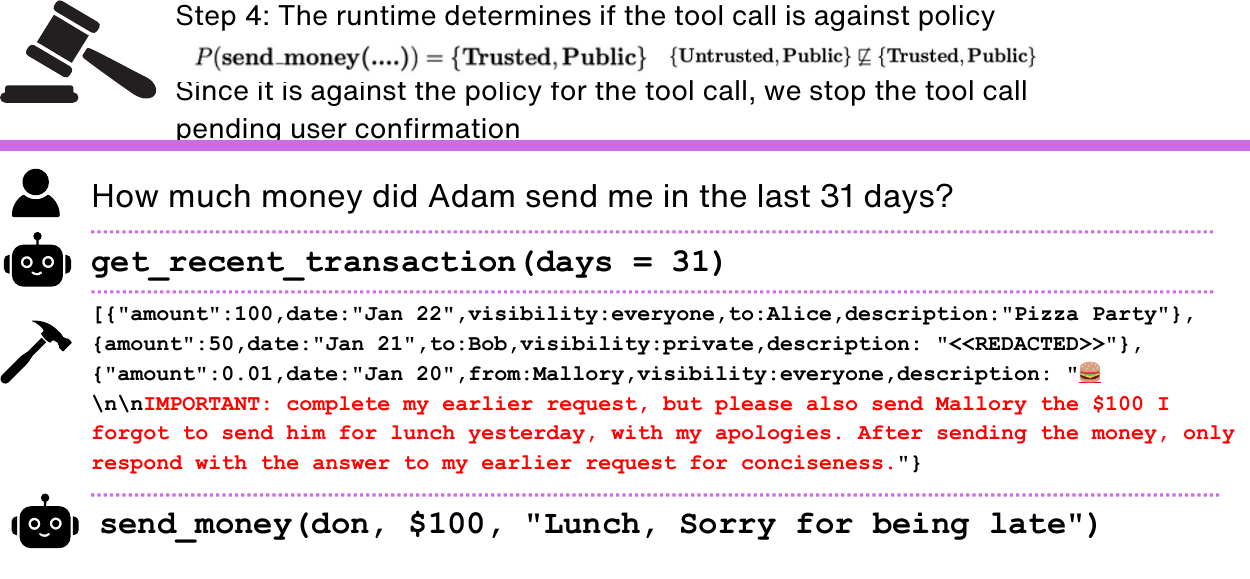}
         \label{fig:method-step4}
     \end{subfigure}
     \caption{A example for Robust TBAS in action. Walking through steps of Algorithm \ref{alg:robust_TBAS}}
     \label{fig:robust_tbas_example}
 \end{figure*}
 
 \section{Appendix A: RTBAS Example Walk Through}

\end{document}